%% file: main.tex
\documentclass[a4paper]{article}
\usepackage[utf8]{inputenc}
\usepackage{graphicx}
\usepackage{amsmath}
\usepackage{color, colortbl}
\usepackage{booktabs}
\usepackage{mathtools}
\usepackage{enumitem} 
\usepackage[affil-it]{authblk} 
\usepackage{multirow}
\usepackage{diagbox}
\usepackage{subfig}
\usepackage{hyperref} 
\usepackage{listings}
\usepackage{floatpag}
\usepackage{float}
\usepackage[labelfont=bf,figurename=Fig.]{caption}
\usepackage{array}
\usepackage{csquotes}

\usepackage[style=authoryear-ibid,backend=biber]{biblatex}
\usepackage{fullpage}

\newcolumntype{$}{>{\global\let\currentrowstyle\relax}}
\newcolumntype{^}{>{\currentrowstyle}}
\newcommand{\rowstyle}[1]{\gdef\currentrowstyle{#1}%
  #1\ignorespaces
}
\title{Objective hearing threshold identification from auditory brainstem response measurements using supervised and self-supervised approaches}

\author[1,3,$\dag$]{Dominik Thalmeier}
\author[2,$\dag$]{Gregor Miller}
\author[2]{Elida Schneltzer}
\author[2]{Anja Hurt}
\author[2,6,7,$\ddag$]{Martin Hrab\v{e} de Angelis}
\author[2]{Lore Becker}
\author[1,3,4,5,$\ddag$]{Christian L. Müller}
\author[2]{Holger Maier}

\affil[1]{Institute of Computational Biology, Helmholtz Zentrum München}
\affil[2]{Institute of Experimental Genetics, Helmholtz Zentrum München}
\affil[3]{Helmholtz AI, Helmholtz Zentrum München}
\affil[4]{Department of Statistics, LMU München, München}
\affil[5]{Center for Computational Mathematics, Flatiron Institute, New York}
\affil[6]{German Center for Diabetes Research (DZD), Neuherberg, Germany}
\affil[7]{Chair of Experimental Genetics, School of Life Science Weihenstephan, 
Technische Universität München, Freising, Germany}

\affil[$\dag$]{first authors}
\affil[$\ddag$]{corresponding authors}

\date{}

\begin{document}
\maketitle

\begin{abstract}
Hearing loss is a major health problem and psychological burden in humans. Mouse models offer a possibility to elucidate genes involved in the underlying developmental and pathophysiological mechanisms of hearing impairment. To this end, large-scale mouse phenotyping programs include auditory phenotyping of single-gene knockout mouse lines. Using the auditory brainstem response (ABR) procedure, the German Mouse Clinic and similar facilities worldwide have produced large, uniform data sets of averaged ABR raw data of mutant and wildtype mice.

In the course of standard ABR analysis, hearing thresholds are assessed visually by trained staff from series of signal curves of increasing sound pressure level. This is time-consuming and prone to be biased by the reader as well as the graphical display quality and scale. 
 
In an attempt to reduce workload and improve quality and reproducibility, we developed and compared two methods for automated hearing threshold identification from averaged ABR raw data: a supervised approach involving two combined neural networks trained on human-generated labels and a self-supervised approach, which exploits the signal power spectrum and combines random forest sound level estimation with a piece-wise curve fitting algorithm for threshold finding.

We show that both models work well, outperform human threshold detection, and are suitable for fast, reliable, and unbiased hearing threshold detection and quality control. In a high-throughput mouse phenotyping environment, both methods perform well as part of an automated end-to-end screening pipeline to detect candidate genes for hearing involvement. 
Code for both models as well as data used for this work are freely available.
\end{abstract}

\section{Introduction}

Impaired hearing has a high impact on quality of life and age-related hearing loss is a common health burden in an aging society \autocite{james2018,cunningham2017}. Disease models of hearing loss using mutant mouse lines can be useful for research of the underlying pathophysiological and molecular mechanisms. Using auditory brainstem response (ABR), a large-scale screen of 1,211 single-gene knock-out mouse lines has recently identified dozens of candidate genes associated with hearing threshold impairment \autocite{ingham2019}. Earlier, an even larger study \autocite{bowl2017} revealed 52 novel candidate genes with hearing loss involvement from analysis of ABR data measured on 3,006 mutant mouse strains within the International Mouse Phenotyping Consortium (IMPC) \autocite{dickinson2016, meehan2017} effort.

The ABR is a form of electroencephalography (EEG), in which electrical potentials are recorded from the scalp of clinical patients or laboratory animals as evoked responses to auditory stimulation. Response signals following rapidly repeated stimulus sequences are averaged and produce typical ABR waveforms that are characterised by specific peaks and troughs, their amplitudes and latencies. As the ABR results from neurological and involuntary processing of sound signals by the different regions of the auditory brainstem, it is an easy-to-perform diagnostic method for hearing assessment of unconscious patients, infants, or animals. In large-scale mouse phenotyping and - more generally - in basic auditory research, it is established as standardised method for measuring hearing function for many years \autocite{ingham2011}. 

When using ABR for hearing threshold identification, this involves a series of measurements at increasing sound pressure levels (SPL) in 5 dB steps at different pure-tone frequencies (``tone pips'') at 6, 12, 18, 24, and 30 kHz as well as a broadband frequency stimulus ``click''). For each tone pip and click stimulus, the ABR waveforms are displayed in a stacked diagram ordered by ascending SPL. In this audiogram, the hearing threshold (HT) for each frequency is then determined as the lowest SPL where a trained human reader can still detect a signal during a visual assessment of the stacked curves diagram. This signal has to be consistent with higher SPL signals, i.e. exhibiting the same, however weaker and shifted peaks. A plot of hearing threshold SPLs vs. stimulus frequency ``hearing curve'') allows rapid overall characterisation of hearing sensitivity.

It is well-established that threshold determination by human readers is prone to reader bias \autocite{gans1992} as well as intra- and inter-reader variability \autocite{arnold1985, vidler2004, zaitoun2014}. This might depend on different visualisation tools, reader concentration, experience, training, and personal visual skills.  In particular in high-throughput environments, maintaining the same conditions over hours is difficult. Another challenge is to achieve and maintain low inter-reader variability in teams with different readers. 

Accordingly, since early on in ABR application, there have been attempts to automate and develop objective methods to determine hearing thresholds from ABR measurements. Over the years, ABR has been discussed in literature as \textit{Auditory Evoked Potentials} (AEP) \autocite{paulraj2014}, \textit{Cortical Auditory Evoked Potentials} (CAEP) \autocite{carter2010}, \textit{Brainstem auditory evoked potential} (BAEP) \autocite{alpsan1994, vannier2002}, \textit{Brainstem Evoked Response Audiometry} (BERA) \autocite{lundt2019}, and \textit{Auditory Evoked Potential} (EAP) \autocite{dobie1989}. Many approaches applied and combined methods from different fields of statistics \autocite{acir2006, arnold1985, berninger2014, bogaerts2009, carter2010, cebulla2000, chesnaye2018, cone1997, dobie1989, dobrowolski2016, mccullagh2007, ozdamar1990, ozdamar1994, schilling2019, suthakar2019, wang2020}, often involving feature extraction from the time and/or the frequency domain. Some approaches also involved bootstrapping \autocite{lv2007}, comparison to templates \autocite{cone1997, vannier2002}, or deep learning \autocite{alpsan1994, davey2007, paulraj2014, mckearney2019, chen2021}.

While most of the published methods for automated threshold identification use averaged response data, a recently published method \autocite{wang2020} processes individual sweep responses with good results. Unfortunately, although always generated during ABR, individual sweep response time curves are not always easily accessible. Instead, readers are usually only provided with the averaged curves. 

Despite all these published efforts, automated approaches seem to have not yet replaced the visual threshold identification by experienced human readers in research practice. This is unfortunate since determining hearing thresholds in thousands of mice is not only laborious and subjective, as discussed above. In addition, long-term structured phenotyping efforts as performed in the German Mouse Clinic (GMC) \autocite{gailus-durner2005, fuchs2009} or in the IMPC generate a huge corpus of ABR data. When it comes to big data analysis, ensuring objective, accurate, and same-standard threshold reading across the whole data set is hardly feasible with human readers.

In this work, we present our efforts and results towards developing a solution for objective and automated high-throughput identification of hearing thresholds from averaged ABR raw data in large-scale research environments. It is intended to reduce human workload, generate accurate, objective, and reproducible results, re-evaluate legacy data, and establish automated quality control processes.

Using a data set generated at the German Mouse Clinic within the IMPC effort as well as an independent external data set provided by the Wellcome Sanger Institute, we developed both a supervised and a self-supervised automated threshold detection method that work on the averaged data available to the researcher. Performance and quality of both methods are compared to the gold standard manual threshold detection method and to each other using two independent data sets. Furthermore, we developed an evaluation method that allows relative comparison of threshold detection methods without requiring any kind of ground truth.

In addition, we developed and evaluated data processing and visualisation methods that allow rapid identification of hearing involvement candidate genes using comparative manual and automated threshold finding.

\section{Materials and methods}

\subsection{Data generation}\label{data_generation}

In this work, averaged ABR raw data from measurements conducted in the German Mouse Clinic on mice from both sexes at fourteen weeks of age was used. The ABR measurements were performed as part of a large-scale, primary comprehensive phenotyping effort within the IMPC.  Accordingly, the data set comprised mutant mice, representing hundreds of different single-gene knockouts, as well as control wildtype mice. All mice were either on a C57BL/6NTac or C57BL/6NCrl genetic background and measured between 2013 and 2020. Original mouse husbandry and animal experiments were carried out in accordance with European Directive 2010/63/EU and following the approval by the responsible local authority of the \textit{Regierung von Oberbayern}, Germany. Mice  were group-housed in standard individually ventilated cages under a 12h light/dark schedule in controlled environmental conditions of $22\pm2$ °C and $50\pm10\%$ relative humidity and fed a normal chow diet and water \textit{ad libitum}. Measurements were performed mainly in the morning.

Mice were anaesthetised with ketamine/xylazine and transferred onto a heating blanket in a sound-attenuating booth. Subcutaneous needle electrodes were inserted in the skin on the vertex (active) and overlying the ventral region of the left (reference) and right (ground) bullae. Stimuli were presented as free-field sounds from a loudspeaker in front of the interaural axis. The sound delivery system was calibrated using a microphone (PCB Piezotronics). For threshold determination, custom software (kindly provided by the Wellcome Sanger Institute) and Tucker Davis Technologies hardware were used to deliver click (0.01 ms duration) and tone pip (6, 12, 18, 24, and 30 kHz of 5 ms duration, 1 ms rise/fall time) stimuli over a range of sound pressure levels (SPL) in 5 dB steps (Click: 0-85 dB, 6 kHz: 20-85 dB, 12-24 kHz: 0-85 dB, 30 kHz: 20-85 dB). Averaged responses to 256 stimuli, presented at 42.6/s, were analysed. For manual threshold detection, the lowest sound intensity giving a visually detectable ABR response was determined. For further reference, this data set is addressed as the \emph{GMC} data set.

To test the methods with external data, a large, published resource of ABR raw data from the Wellcome Sanger Institute \autocite{ingham2019data} measured on 9,000+ mice from 1,211 single-gene mutant lines and respective control (wildtype) mice on largely C57BL/6N but also other genetic backgrounds was used. We thank the authors for kindly making this invaluable resource publicly available. This data set is addressed as the \emph{ING} data set.

\subsection{Data pre-processing}
All ABR data used was pre-processed to create a single csv file containing the ABR time series (columns t0 - t999), an individual mouse identifier, stimulus frequency, stimulus SPL, and a manually determined hearing threshold. For each mouse, there are different ABR time series corresponding to six different sound stimuli: broadband click, 6, 12, 18, 24, and 30 kHz, each of which was measured for a range of sound pressure levels. The exact range of sound levels can vary between the different mice and stimuli, as described above. Mice not having a complete set of data for all six stimuli were excluded during pre-processing of the GMC data set.

\subsection{Data validation} \label{data_validation}
\begin{figure}[h]
\centering
\includegraphics[width=0.4\linewidth]{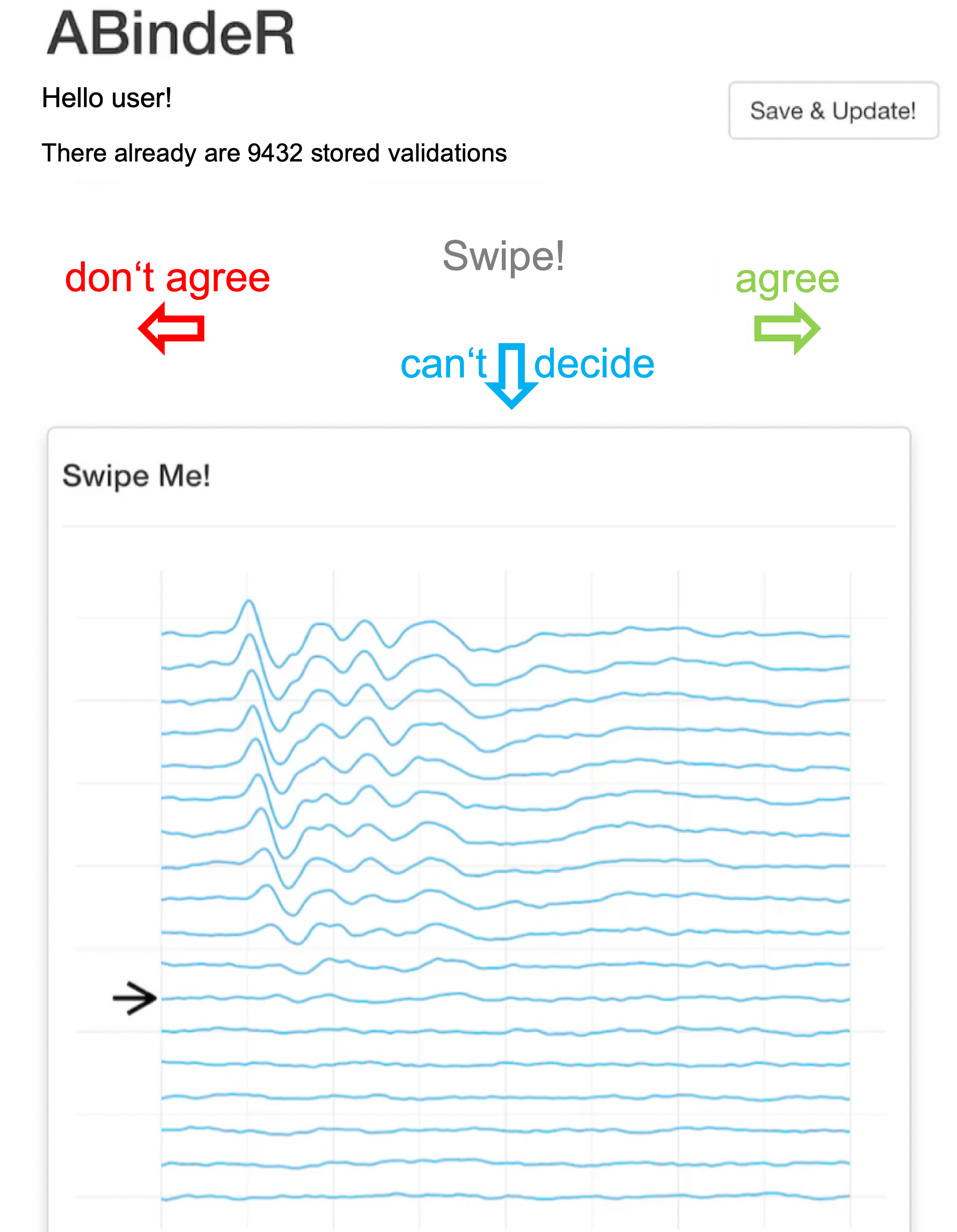}
\caption{\textbf{ABR threshold validation app used on standard tablet computers}. Users are presented a randomly selected ABR measurement along with the original human-assigned threshold, visually indicated by an arrow. Users can enter their evaluation by swiping left: ``don't agree'', right: ``agree'', down: ``can't decide''. No mouse or stimulus information is provided in order to allow unbiased evaluation.}
\label{fig:validationApp}
\end{figure}

In order to obtain the best-possible label quality in the supervised approach, the hearing thresholds of roughly one-seventh of the GMC data set were re-validated using a simple R/shiny app on standard tablet computers, as shown in \mbox{Fig. \ref{fig:validationApp}}. In the app, ABR-trained users had to state their agreement with the original human-assigned threshold for randomly presented hearing curves. Measurements with an ``agree'' validation result were subsequently weighted higher in the supervised neural network approach (see \ref{neuralNetworkApproach}) than the measurements receiving a ``don't agree'' or ``can't decide'' validation result. Using the app, a large number of ABR measurements could be re-evaluated in short time in a blinded fashion, since no information about the mouse, the stimulus, or the original reader is provided whatsoever.

The hearing thresholds of the ING data set were not re-validated, but used as provided. 

\subsection{Supervised artificial neural network (NN)}\label{neuralNetworkApproach}
\begin{figure}[h]
\centering
\includegraphics[width=0.95\linewidth]{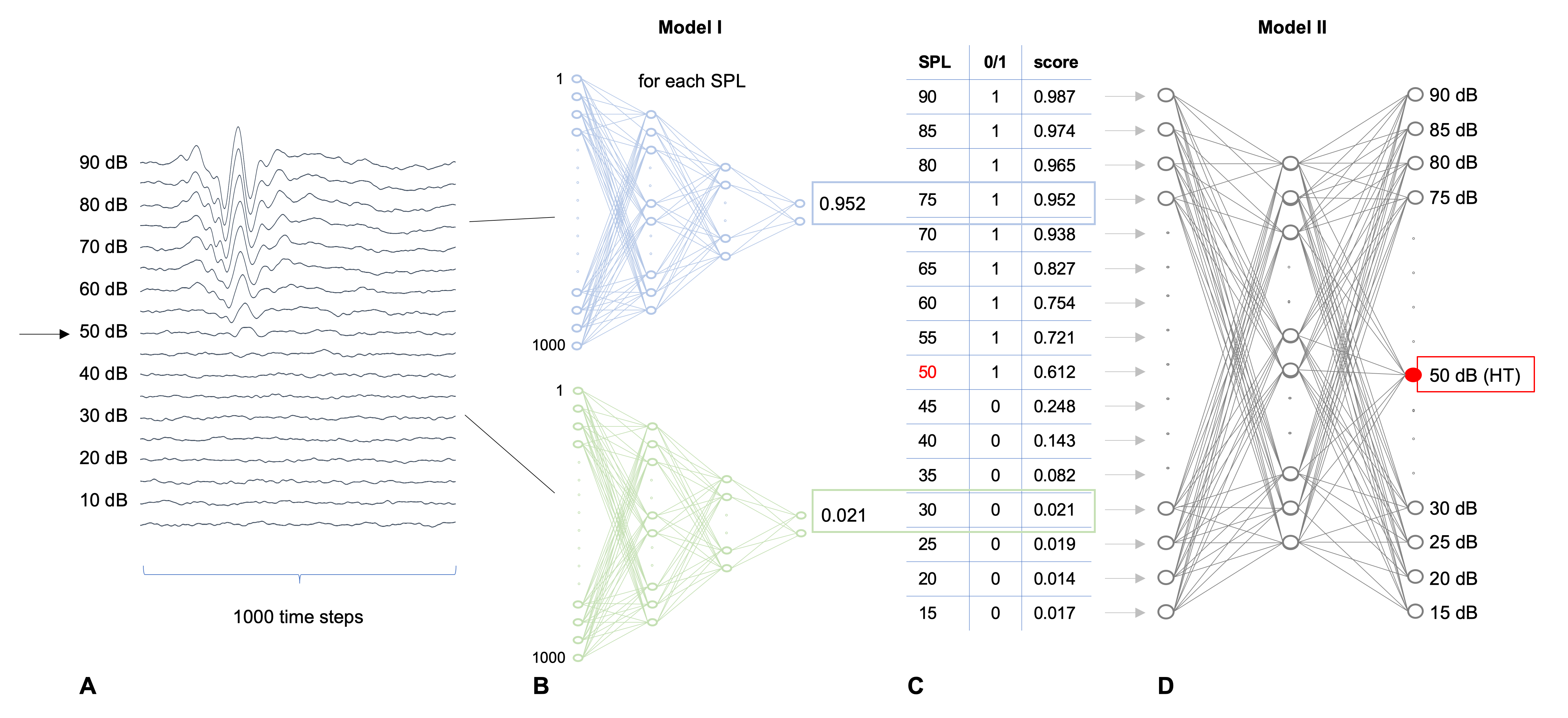}
\caption{\textbf{Scheme illustrating the two-stage method}. \textbf{A} Example ABR plot for one stimulus. Stacked curves correspond to evoked response time signal (1,000 time steps) for increasing sound pressure levels (SPL). The arrow indicates the human-assigned hearing threshold. \textbf{B} For each curve of all sound pressure level (SPL), a trained neural network (Model I) classifier predicts if a response is present (1, blue example) or not (0, green example), using a 1,000 time step input vector and delivering a class score as output. \textbf{C} Result of Model I classifier. For each SPL, the binary decision (0/1) and a class score is generated. \textbf{D} A second classifier (Model II) uses the class score outputs from Model I as input vector and predicts the hearing threshold (HT, red). Both models are trained on the actual hearing threshold label.}
\label{fig:twoStageNN}
\end{figure}

For modelling the human threshold finding process, a two-stage approach was implemented, which is illustrated in \mbox{Fig. \ref{fig:twoStageNN}}. A convolutional neural network (Model I) is trained as classifier to predict if an ABR response is present or not present in a single stimulus curve (one frequency, one sound pressure level). The required labels for Model I are derived from the original hearing thresholds under the assumption that all sub-threshold SPL curves represent non-hearing, while threshold and supra-threshold SPL curves represent hearing. A second convolutional neural network (Model II) is then trained as classifier for every stimulus to predict the hearing threshold using the respective class score outputs of Model I as input and the original hearing thresholds as labels. A five-fold grouped cross-validation approach with the mice as groups was followed. First, mice were randomly split 4:1 into training and test mice. This training data was then randomly split 4:1 into training and validation mice in each fold. The architecture of both models is provided in \nameref{sup1:model_architectures}. For reference, this method will be addressed as ``NN'' in this work.

\subsection{Self-supervised Sound Level Regression (SLR)} \label{slr_method}
\begin{figure}[h]
\centering
\includegraphics[width=0.9\linewidth]{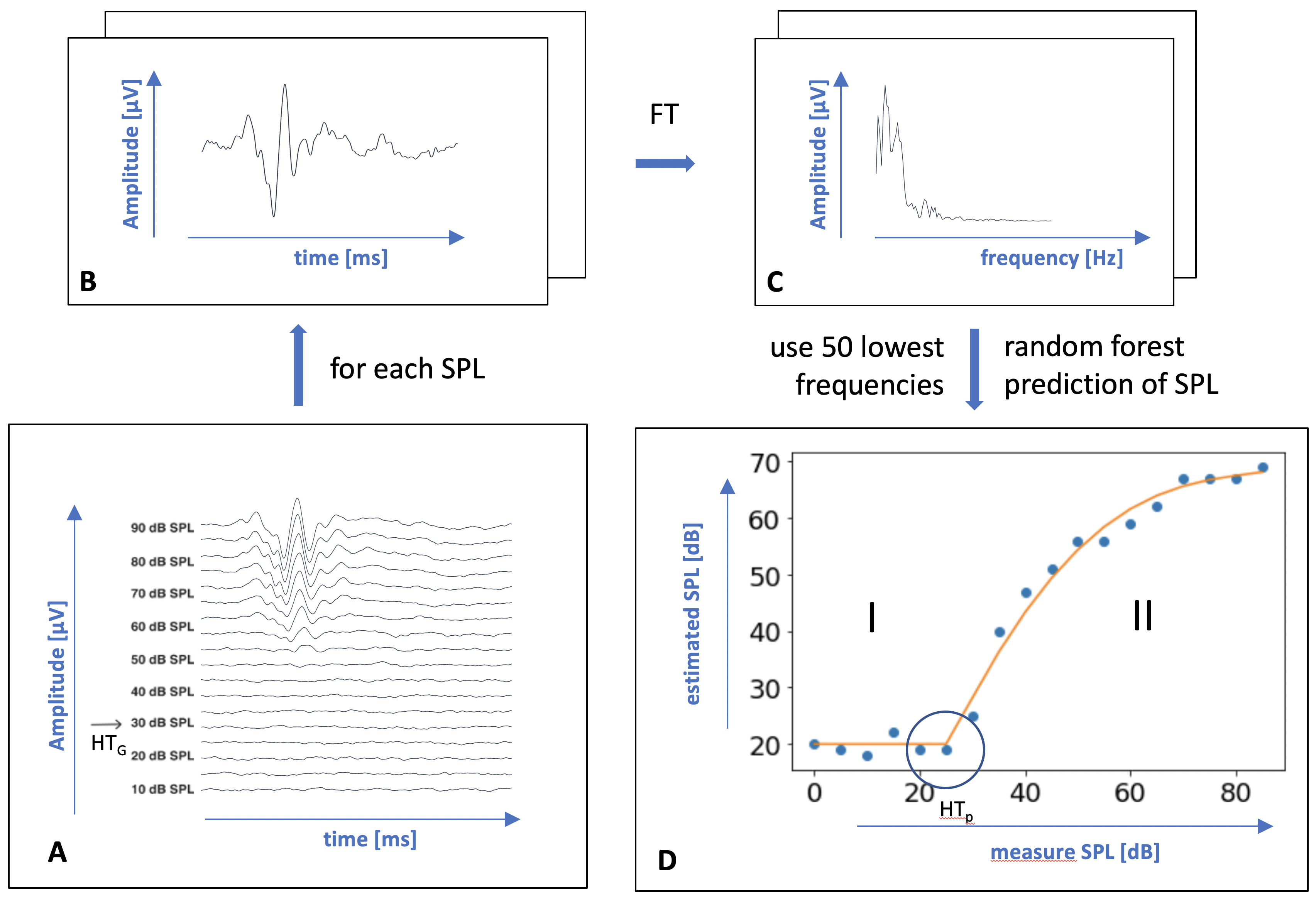}
\caption{\textbf{Scheme illustrating self-supervised prediction of hearing thresholds from evoked ABR signals}. \textbf{A} Stacked ABR time series for different given sound pressure levels (SPL) at a given stimulation frequency, e.g. 18 kHz. The arrow indicates the manual hearing threshold HT\textsubscript{G}, which is only used for validation in this approach. \textbf{B} Amplitude signal in the time domain for a single SPL. \textbf{C} Amplitude in the frequency domain after Fourier transformation (FT). The lowest 50 frequencies are extracted from the power spectrum as features. \textbf{D} estimated vs. measure SPL. Blue dots: sound pressure levels estimated by random forest regression plotted against the actual measure sound levels. Orange line: a piece-wise function composed of a constant (I) and a polynomial (II) part. The polynomial part of the function is fitted using an elastic net. The circle indicates where the constant and polynomial function meet, which determines the predicted hearing threshold HT\textsubscript{P}.}
\label{fig:scheme}
\end{figure}

A scheme of the new threshold detection method called ``Sound Level Regression'' is shown in \mbox{Fig. \ref{fig:scheme}}. For reference, this method will be addressed as ``SLR'' in this work. In short, it consists of two steps, which are performed on each stimulus frequency and click separately:

\begin{enumerate}[label=\Alph*]
    \item \textbf{Sound level estimation from single curves} \\
    In this step, the sound level of the stimulus is estimated from the time series data of its evoked signal curve using a supervised regression method. More precisely, as the sound level is given in the data itself, it is called a self-supervised method. The core idea is that such a prediction can only work if the sound level of the stimulus that leads to the evoked signal curve is above the hearing threshold. As otherwise, per definition, no information about the sound level should be contained in the resulting time series. 

    \item \textbf{Hearing threshold estimation from sound level estimates} \\
    In this step, the hearing threshold can be extracted from the predicted sound levels. It can be expected that for sub-threshold conditions, the predicted sound levels fluctuate around a constant value, while for supra-threshold conditions, the predicted sound levels follow a monotonically increasing function of the actual sound level (see \mbox{figure \ref{fig:scheme}D}). By fitting a piece-wise function that is constant up to a certain value and then monotonically increasing, the break point can be used as an estimate of the hearing threshold.

\end{enumerate}

In the following, the two steps are described in more detail.

\subsubsection{Step A: Estimate sound levels for hearing curves using machine learning}

In order to estimate the sound levels, the time series are first transformed into a feature space. Using Fourier transformation (FT), the power spectrum of the signal is computed. Due to the sparsity of this representation, only the lowest 50 frequencies are used as features. Then, a random forest regression model is trained to estimate the sound level for each hearing curve from the corresponding feature vector.

To avoid overfitting, training and prediction are embedded in a 5-fold grouped cross-validation with the mice as groups. In each fold, mice are divided into a training and a test group. The random forest is trained only on time series from the training group and makes the prediction for the test group. This way, training is strictly separated from the test data and prediction is still possible for each time series.

\subsubsection{Step B: Determine hearing thresholds from sound level estimates}

Next, the predicted sound levels are used to determine the hearing threshold. As described above, a piece-wise function is fitted, consisting of a constant part and a monotonically increasing part, which is modeled as a polynomial function. In principle, the breakpoint of the fitted function could be used directly to determine the hearing threshold. However, we have found that this is not very robust, since it is possible that the polynomial starts as a very flat function that is still quite similar to a constant function. Therefore, the hearing threshold is determined as the sound level at which the polynomial part of the fitted function deviates from the constant part by more than 4 dB. In the remainder of this section, details about the fitting process are described.

\paragraph*{Determine upper and lower bounds for threshold}
First, the search space for the hearing threshold is narrowed by calculating a rough estimate of its upper and lower bounds. 

The upper bound of the threshold is determined by the largest sound level for which all estimated values above that limit show a significant positive correlation to the actual sound level used. This is calculated by testing the hypothesis for each sound level in question to see if the sound levels greater than that level have a positive correlation with the corresponding predicted sound levels. The largest value for which the p-value is greater than 5 percent after a Bonferroni correction is used.

As the lower limit for the threshold is determined by the first increase of a function learned by isotonic regression, which empirically was found to be a conservative lower limit for the hearing threshold.

\paragraph*{Fitting a piece-wise function}
What remains is a range between these upper and lower thresholds as candidates for the threshold. Since measurements are taken in steps of 5 dB, possible candidates for the threshold are also limited to a grid of 5 dB. 

For each possible threshold value, a piece-wise function with the breakpoint at the possible threshold position is fitted. The function consists of a constant function on the left side of the breakpoint and a polynomial of the fourth degree for sound levels larger than the breakpoint. An elastic net with l1 ratios of 0.5 and 0.99 and 5-fold cross-validation with automatic determination of the regularisation parameter is used for fitting. Of the various functions used for fitting, one for each possible breakpoint, the one that has the least cross-validation error is selected.

With this procedure it can happen that the true threshold value is e.g. 25.1 dB and therefore a threshold value of 25 is estimated. However, the threshold should be the lowest recorded sound level at which the mouse exhibits stimulus-induced ABR activity, which in this case would be 30 dB.

Therefore, also the piece-wise function for sound levels that are 0.5 dB lower and higher than the selected breakpoint is fitted. If either of these show a cross-validation error that is lower than the current optimum breakpoint, the new value is considered the new optimum and therefore the final predicted hearing threshold.


\subsection{Evaluation curves}
To avoid the use of human-derived ground truth labels in the quality assessment of two hearing threshold finding methods, evaluation curves were developed as a visual quality assessment method. This section describes the theoretical concept behind it.

Assuming that the true hearing thresholds are known, the sample average of all super-threshold curves and the sample average of all sub-threshold curves can be calculated. When taking the sample average of all super-threshold curves, a temporal pattern should emerge, since the mice react to the signal tone in a temporally coherent manner. In contrast, averaging the sub-threshold ABR curves should result in a constant signal as the ABR curves are/have to be temporally incoherent due to the absence of a perceived signal and therefore any temporal pattern is averaged out when taking the mean.

From this argumentation, measures to assess the quality of any threshold finding method can be derived. To this end, all ABR curves from all mice that correspond to a specific stimulus (e.g. click) are given an index $i\in [0,N]$, with $N$ being the total number of measured ABR curves for all mice, but restricted to this stimulus.

Now 
$$\text{l}(i) \coloneqq \frac{\text{soundlevel}(i)}{\text{threshold}(i)}$$
is defined as the threshold normalized sound level. The ABR curves are sorted by $\text{l}(i)$, so that 
\mbox{$\text{l}(i)<\text{l}(i+1)$}. Let $x_i(t)$ with $t \in [0,T]$ be the time series of the ABR curve with index i.

The cumulative average for the first $n$ curves with the lowest threshold normalized sound levels can be computed as 
$$ \bar{X}_n(t) \coloneqq \frac{1}{n}\sum_{i=0}^n x_i(t),$$
where $x_i(t)$ is the time series of the ABR curve with index i defined in the measurement interval $t \in [0,T]$.
Now let $i_\text{crit}$ be the largest index for curves which are still below the threshold value, i.e. for which ${\text{l}(i> i_\text{crit})\geq 1}$ and ${\text{l}(i<i_\text{crit})< 1}$.
Then  for $n \leq i_\text{crit}$, $\bar{X}_n(t)$ should be an approximately\footnote{The 'approximately' is due to the finite sample size.} constant signal with a vanishing temporal variance
$$ S^2(n) \coloneqq \frac{1}{T}\int_0^T {\bar{X}_n(t)}^2 dt - \left(\frac{1}{T}\int_0^T {\bar{X}_n(t)} dt\right)^2\approx 0.$$

If ground truth threshold is used for this sorting, the averaged curve should not deviate significantly from a constant signal until all sub-threshold curves have been added to the cumulative mean $\bar{X}_n(t)$. 
However, if suboptimal threshold values are used, the averaged signal should start to deviate from a constant signal earlier, because true sub-threshold curves are mixed with super-threshold curves.

As an example, there might be a total of $i_\text{crit} = $3000 real sub-threshold curves in the set of all curves. Then $S^2(n) \approx 0$ for $n<=3000$, given the true thresholds are used for sorting. However, if erroneous thresholds are used for sorting, then $S^2(3000)$ can only be zero if the error of the thresholds is a systematic and constant shift of the thresholds. However, if the error is due to an inconsistency in the threshold labeling, then $S^2(3000)>0$, since lower and upper threshold curves are mixed.

Based on this, evaluation curves can be constructed that compare the quality of threshold value procedures: the (normalized) time variance of the averaged signal 
$$\frac{S^2(n)}{S^2(N)}$$
is plotted versus $\frac{n}{N}$, the total percentage of ABR curves included in the cumulative average. 

For the ground truth threshold, this curve should be approximately zero until $\frac{n}{N}$ is equal to the number of sub-threshold curves divided by the total number of ABR curves (= sub + super threshold). After that it should increase.
For suboptimal thresholds, the curve should start to deviate from zero already at smaller levels of $\frac{n}{N}$. The more error-prone the threshold values are, the faster the corresponding evaluation curve deviates from zero.

\section{Results and Discussion}

\subsection{Pre-processing and characterisation of working data sets}
Following pre-processing and validation of raw data, two independent working data sets were produced as described in \ref{data_generation}. In short, the GMC data set is based on \textit{in-house} data collected at the German Mouse Clinic, whereas the ING data set is based on a large published ABR data ressource. \mbox{Table \ref{tab:twoDatasets}} summarises basic properties of the two data sets. 

\input{tables/twoDatasets}

They comprise data of a combined total of 12,391 mice, of which 8,784 \mbox{(2,654 + 6,130)} are mutants and 3,607 \mbox{(1,707 + 1,900)} are controls. In the GMC data set, male and female mice are represented equally, both in the mutant and the control groups. For the ING data set, no information about sex is given. The number of knockout genes represented in the GMC and ING data set is 352 and 1,152, respectively. Twelve genes (\textit{Bach2, Cdkal1, Dbn1, Dnase1l2, Entpd1, Gsk3a, Hdac1, Klk5, Nxn, Rnf10, Slc20a2, Ubash3a}) are common to both sets, resulting in a combined total of 1,492 \mbox{(352 + 1,152 - 12)} knockout genes. The median size of mutant cohorts in the GMC data set is 8, compared to 4 in the ING data set.

\begin{figure}[H]
\centering
\includegraphics[width=0.9\linewidth]{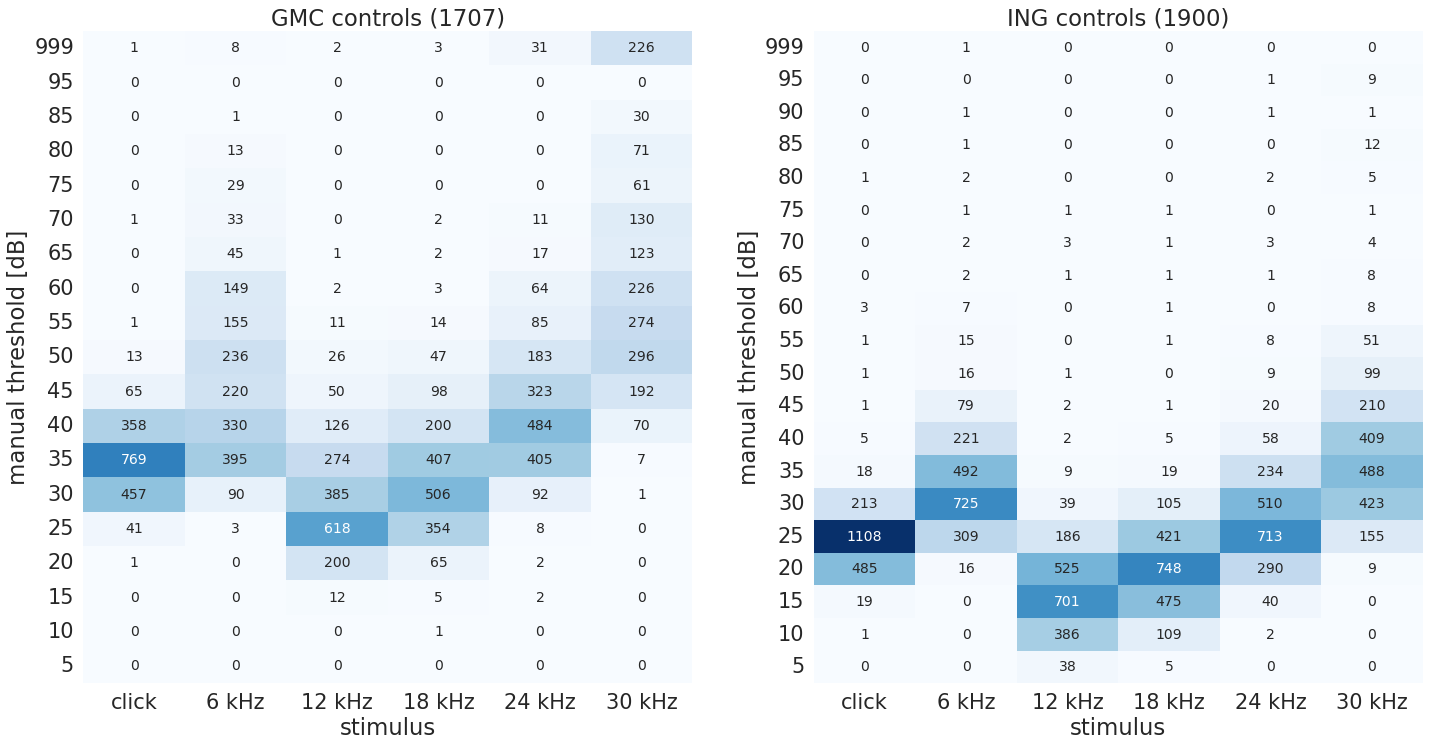}
\caption{\textbf{Available threshold labels for wildtype (control) animals in the GMC (left) and the ING (right) data set.} Numbers in parentheses denote the size of the respective data set. Columns indicate the stimulus (click, 6, 12, 18, 24, 30 kHz), rows indicate the hearing threshold assigned by human readers. Non-hearing was arbitrarily assigned 999. Numbers in cells correspond to the distinct number of animals exhibiting the respective hearing threshold at the given stimulus. To facilitate visual comparison of the data sets, numbers in cells are colour-coded in shades of blue. }
\label{fig:comparisonAvailableLabels}
\end{figure}

To investigate the distribution of human-assigned hearing thresholds in the data sets, the according numbers of control (wildtype) mice have been compiled from raw data and visualised in \mbox{Fig. \ref{fig:comparisonAvailableLabels}}. While the pattern of the hearing threshold labels reflects the typical U-shaped appearance of a hearing curve, it is obvious that there is a 10-15 dB shift towards lower thresholds in the ING data set compared to the GMC data set. Also, threshold variance is smaller for the ING data set. Notably, there is a considerable number of ``non-hearing`` (999) labels in the GMC data for 24 kHz and 30 kHz, whereas this is not the case for the ING data. Naturally, the distribution of hearing thresholds is not uniform, i.e. most mice exhibit a hearing threshold only in a small frequent-specific range. Evidently, for any supervised approach, this means that for non-normal thresholds, there are almost no training cases.

Overall, considerable numbers of same-standard, quality-controlled ABR raw data, including metadata and human-assigned threshold labels, have been compiled into two working data sets for further use.

\subsection{Evaluation and comparison of two new threshold finding methods}
In order to comprehensively examine and compare the performance of the two threshold finding methods introduced in this work, a scheme of eight experiments was conceived as shown in \mbox{table \ref{tab:8experiments}}. First, both methods were evaluated in a way that the neural network based method and the Sound Level Regression were tested on subsets of mice from the same data set used for training and calibration, respectively. In a next step, the robustness of both methods was evaluated, to find out to what extent a method trained/calibrated on the GMC data set can be applied on the ING data set and vice versa. 

For all experiments, data set specific labels assigned by human readers were used to calculate accuracy as a quality measure. To take into account that hearing thresholds a) were assigned with a granularity of only \mbox{5 dB} and b) human threshold finding is prone to variability, accuracies were calculated using three match levels - ``exact'': requiring an exact match of label and predicted/estimated threshold, \mbox{``$\pm5$ dB''} and \mbox{``$\pm10$ dB''}: allowing 5 dB and 10 dB mismatch between label and predicted/estimated threshold to still be considered accurate.

\input{tables/8experiments}

\subsubsection{The neural network model (NN) can objectively predict hearing thresholds from averaged ABR raw data}

With each of both data sets, the NN models were trained and tested with subsets of mutant and control mice from the same data set. This corresponds to experiment 1: ``NN GMC-GMC'' and experiment 4: ``NN ING-ING'' as introduced in \mbox{table \ref{tab:8experiments}}.

Five-fold cross-validation showed that the method is robust and predictions can be generalized to the whole data set (not shown). Accuracies calculated for three match levels (see \mbox{table \ref{tab:nn_accuracies}}) show that requiring exact match is not fit for practical use. However, allowing 5 dB and 10 dB mismatch achieves reasonable overall accuracies. This is not surprising, as  labels are assigned by human readers and human threshold variance is well-established in literature \autocite{arnold1985, vidler2004, zaitoun2014} and confirmed by own evaluation experiments with GMC data (see \ref{data_validation}, data not shown). In general, accuracies are highest for the click stimulus. For both mismatch levels, ING accuracies are higher. This may be due to the observed lower label variability in the ING data set, which hints on more consistent label reading.

\input{tables/nn_accuracies}

\begin{figure}[H]
\centering
\subfloat[NN GMC-GMC]{%
  \includegraphics[width=0.48\linewidth]{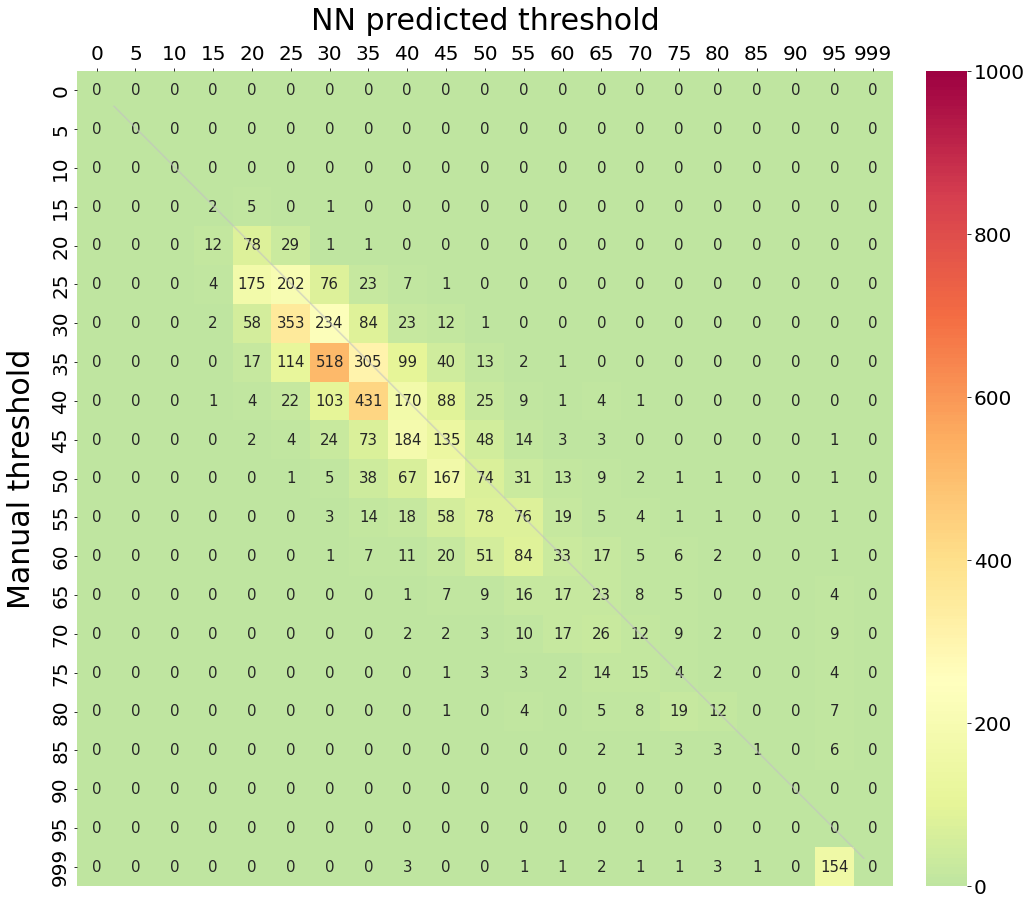}
  \label{fig:confusion1_nn_gmc-gmc}
  }
\subfloat[NN ING-ING]{%
  \includegraphics[width=0.48\linewidth]{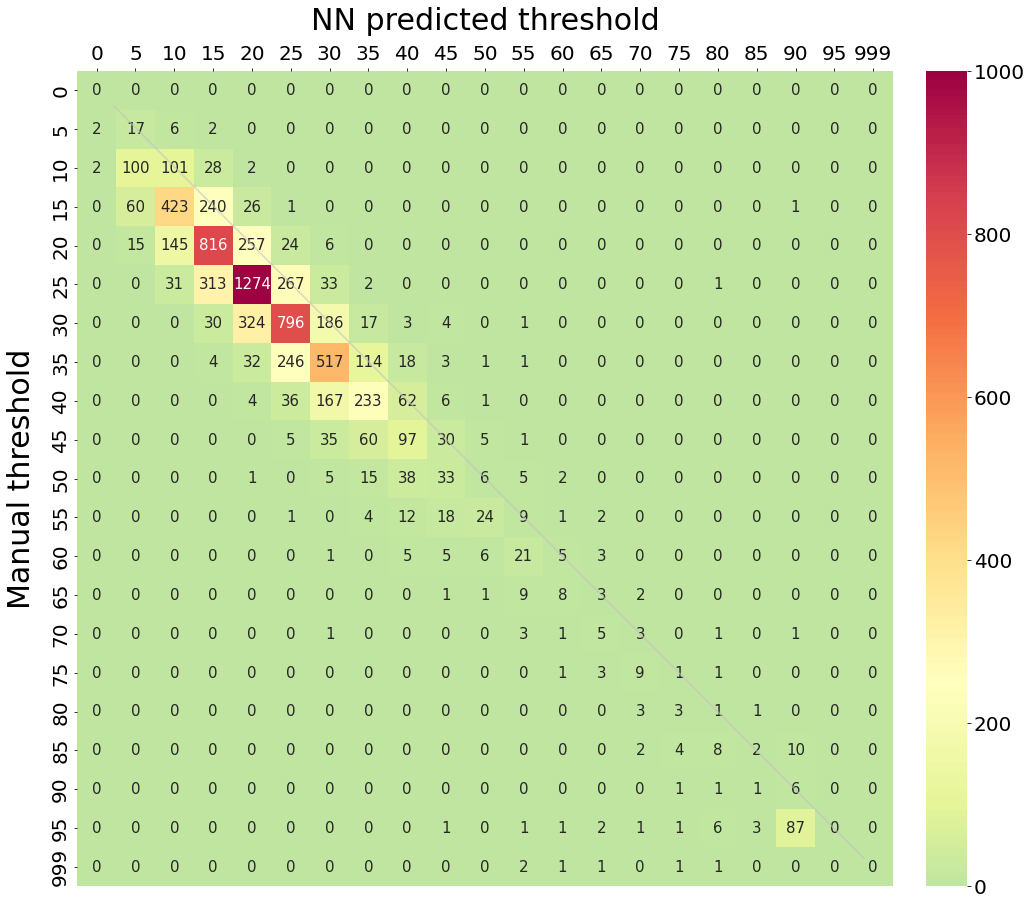}
  \label{fig:confusion4_nn_ing-ing}
  }
\caption{\textbf{Confusion matrix of manual vs. NN predicted hearing thresholds}. In both experiments, named after the scheme introduced in table \ref{tab:8experiments}, the two-stage neural network was trained and tested on subsets of the same data set (left: GMC, right: ING). The threshold manually assigned by human readers is given on the y-axis, the NN predicted thresholds are given on the x-axis. Numbers in cells are numbers of cases across all stimuli. To facilitate interpretation, numbers are coded by colour intensity according to the colour bar. The thin bisecting line indicates an ideal prediction.}
\label{fig:confusion_nn}
\end{figure}

An overall comparison of manual vs. NN predicted thresholds is given in \mbox{Fig. \ref{fig:confusion_nn}} for both experiments. Interestingly, both experiments reveal a \mbox{5 dB} shift towards lower predicted thresholds. However, since manual thresholds are used as labels, but do not necessarily provide a ground truth for the hearing threshold, the question remains whether this difference is due to an inaccuracy in manual thresholds or algorithmic prediction.

\subsubsection{Sound Level Regression (SLR) can objectively predict hearing thresholds from averaged ABR raw data}

With both data sets, the SLR models were calibrated and tested with subsets of mutant and control mice from the same data set. This corresponds to experiment 5: ``SLR GMC-GMC'' and experiment 8: ``SLR ING-ING'' as introduced in \mbox{table \ref{tab:8experiments}}.

Quite similar as with the NN approach, SLR accuracies calculated for three match levels (see \mbox{table \ref{tab:slr_accuracies}}) show that exact match accuracies are far below practical applicability. Again, allowing \mbox{5 dB} and \mbox{10 dB} mismatch achieves reasonable accuracies, however lower than with the NN approach. SLR accuracies were consistently highest for the click stimulus. 

\input{tables/slr_accuracies}

\begin{figure}[ht]
\centering
\subfloat[SLR GMC-GMC]{%
  \includegraphics[width=0.48\linewidth]{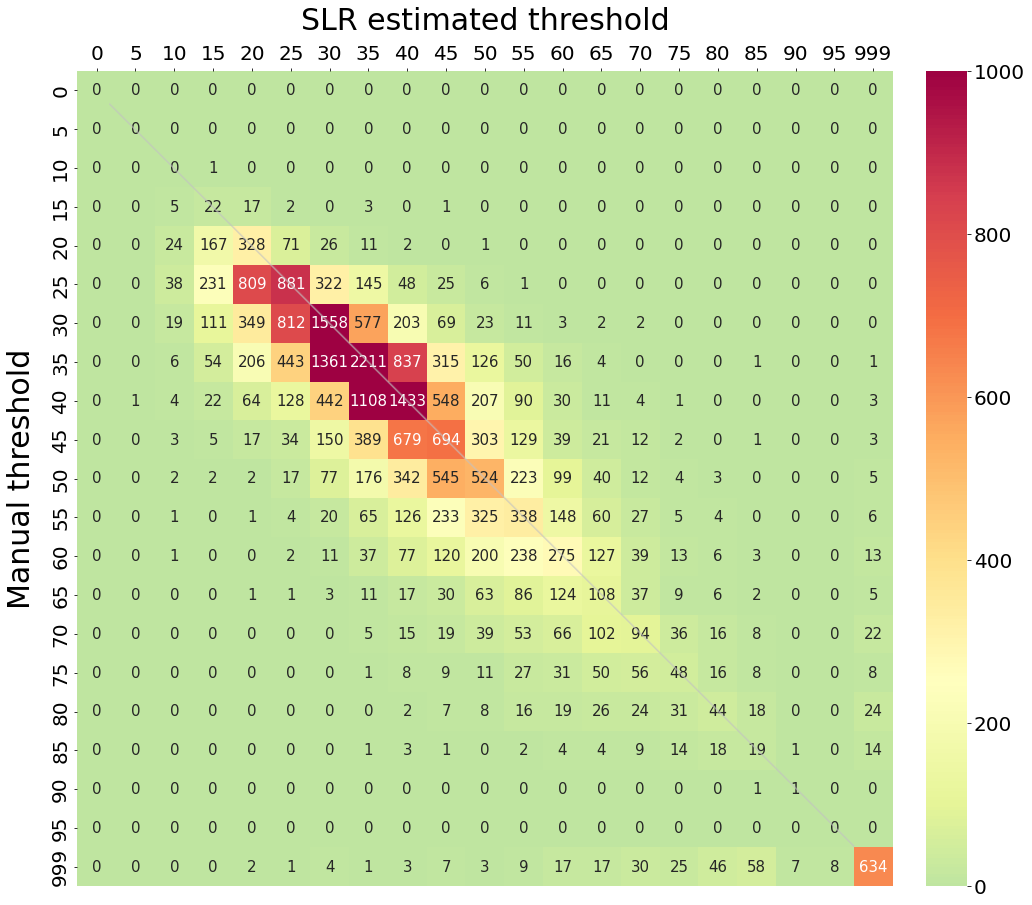}
  \label{fig:confusion5_slr_gmc}
  }
\subfloat[SLR ING-ING]{%
  \includegraphics[width=0.48\linewidth]{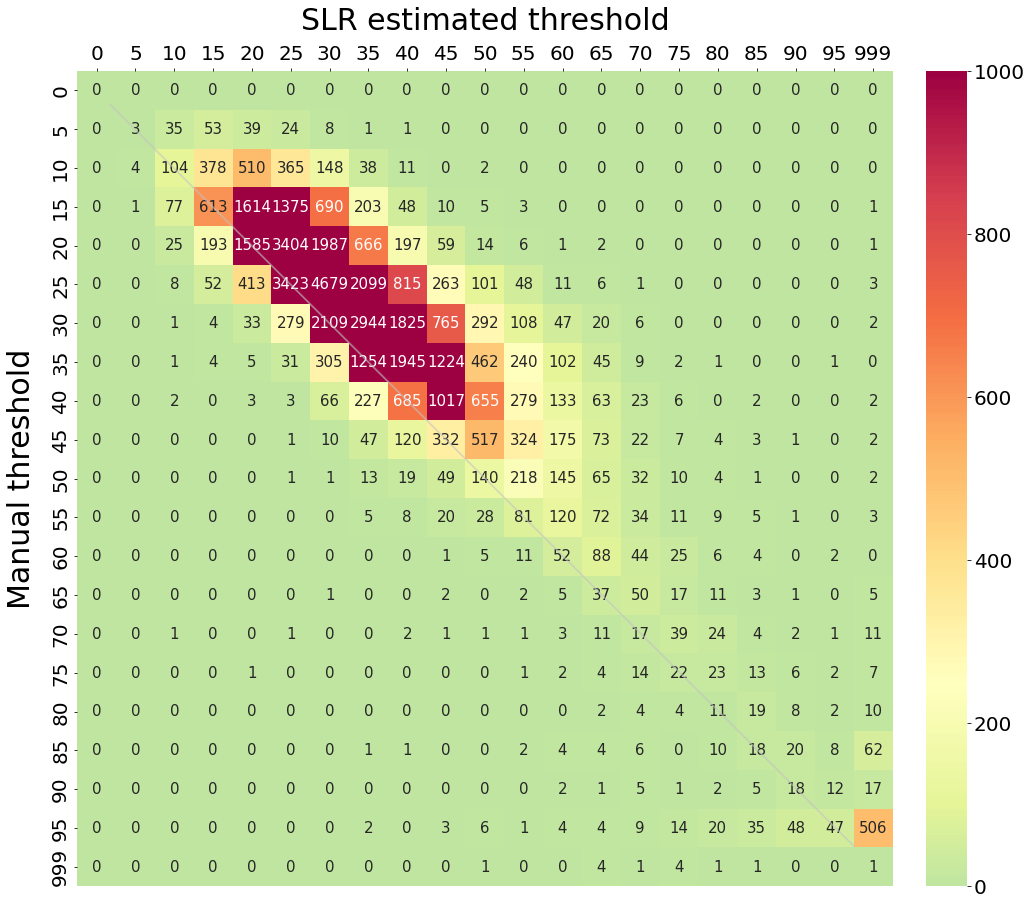}
  \label{fig:confusion8_slr_ing}
  }
\caption{\textbf{Confusion matrix of manual vs. SLR estimated hearing thresholds}. In both experiments, named after the scheme introduced in table \ref{tab:8experiments}, the SLR model was tested on the complete data set after being calibrated on a subset of it (left: GMC, right: ING). The threshold manually assigned by human readers is given on the y-axis, the SLR estimated thresholds are given on the x-axis. Numbers in cells are numbers of cases across all stimuli. To facilitate interpretation, numbers are coded by colour intensity according to the colour bar. The thin bisecting line indicates an ideal estimation. Note: compared to Fig. \ref{fig:confusion_nn}, numbers are much higher, since SLR methodically allows testing with the complete data set.}
\label{fig:confusion_slr}
\end{figure}

An overall comparison of manual vs. SLR estimated thresholds is given in \mbox{Fig. \ref{fig:confusion_slr}} for both experiments. In contrast to the NN approach, the ING experiment reveals a \mbox{5-10 dB} shift towards higher estimated thresholds, while the estimation fits quite well in the GMC experiment. Also in contrast to the NN approach, for both mismatch levels, accuracies are higher for the GMC data set. As the SLR method is independent from human labels, this may hint towards systematic differences in human curve reader training or criteria between the data sets, which is also supported by the visible shift in the manual thresholds (\mbox{Fig. \ref{fig:comparisonAvailableLabels}}).
\\ 
\\
As an overall evaluation of both methods, NN as well as SLR both work well and deliver good results compared to human labels, provided \mbox{5 dB} or even \mbox{10 dB} mismatch are allowed. Depending on the level of reader training and quality control, this may be acceptable to many laboratories. More even so, since both methods have the advantage of delivering reproducible results and are applicable to large ABR data collections while avoiding reader bias.  

\subsubsection{NN shows higher accuracy than SLR, however SLR is more robust}
To investigate robustness of both methods, cross-over experiments according to the scheme laid out in \mbox{table \ref{tab:8experiments}} were performed. In this experiment series, both methods were systematically trained/calibrated on one data set and applied to the other one. For NN, this corresponds to \mbox{experiment 2}: \mbox{``NN GMC-ING''} and \mbox{experiment 3}: \mbox{``NN ING-GMC''}. For SLR, the respective experiments are \mbox{6: ``SLR GMC-ING''} and \mbox{7: ``SLR ING-GMC''}. Resulting accuracies are shown in a large overview \mbox{table \ref{tab:8experiments_accuraries}} for all three match levels.

\input{tables/8experiments_accuraries}

For all experiments, ``exact match'' accuracies are only shown for the sake of completeness and are not further discussed, since they are consistently far below any usability. However, for both \mbox{$\pm5$ dB} and \mbox{$\pm10$ dB} match level, a similar pattern emerges: for both data sets, NN almost always shows highest accuracies when trained on the later test data set (experiments 1 and 4). This is not the case for cross-over experiments 2 and 3, where accuracies collapse. In contrast, SLR exhibits almost consistent accuracies throughout all experiments, thus seems robust and invariant against transfer between calibration and test data sets. This is not surprising, since SLR methodically is not dependent on human labels at all and marginal differences between data sets may only be explained by differences in experimental settings or primary data capture which influence raw data properties. 

Overall, under the condition of having a large amount of high quality human labels delivering a consistent hearing threshold ground truth for training, NN shows to be a good replacement or supplementary method for human threshold reading. SLR could be the method of choice if no or not sufficient consistent human labels are available, since it works purely data-driven and ready calibrated SLR methods can be transferred between data sets.

\subsection{Both NN and SLR can outperform manual threshold finding}
Standard accuracy measurement requires gold standard labels delivering a ground truth. While large, specialised groups may be able to maintain a high level of human reader training and quality control consistently over many years, ABR threshold data generated in smaller groups may show more reader bias and higher variability. Therefore, it seems sensible to measure the quality of any hearing threshold determining method without requiring a gold standard.

Evaluation curves developed in this work are such a method, which has been used to compare human, NN and SLR threshold finding. A forth method always returns a constant threshold, arbitrarily set to 50 dB\footnote{Note that the curve looks the same for any constant threshold.}, and is used as a control, since it can assumed to be the worst method. In short, a method is better than another method, the longer its curve stays closer to zero.

\begin{figure}[h]
\centering
\subfloat[NN/SLR GMC-GMC (experiments 1 and 5)]{%
  \includegraphics[width=0.9\linewidth]{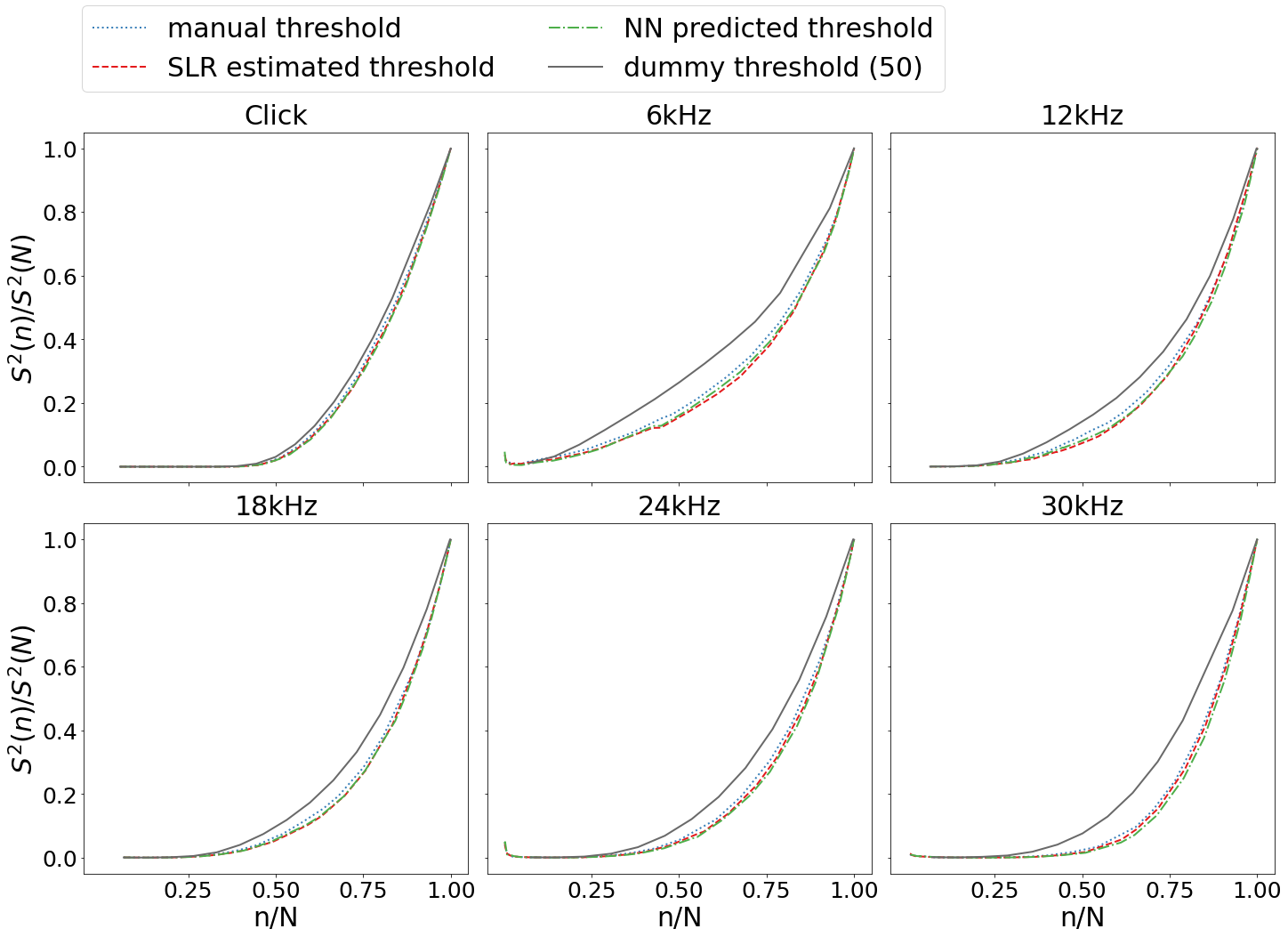}
  \label{fig:evaluation_curves-exp1_5-GMC-GMC}
  }
  
\subfloat[NN/SLR ING-ING (experiments 4 and 8)]{%
  \includegraphics[width=0.9\linewidth]{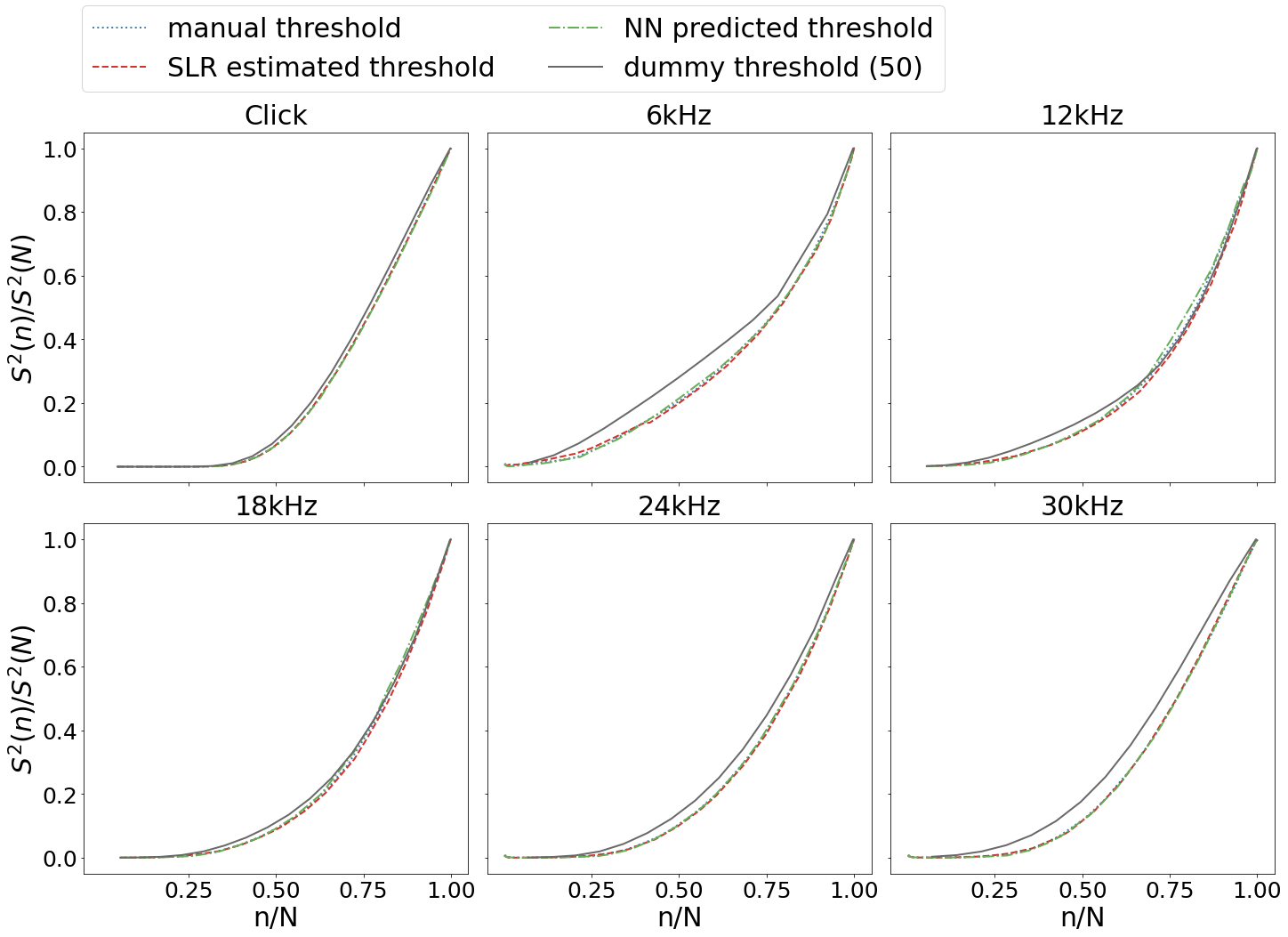}
  \label{fig:evaluation_curves-exp4_8-ING-ING}
  }
\caption{\textbf{Objective comparison of threshold finding methods using evaluation curves}. Four methods are compared: manual thresholds (blue, dotted lines), SLR estimations (red, dashed lines), NN predictions (green, dash-dotted lines), and an ``always 50 dB'' control method (grey, solid lines). Separate plots show evaluation curves for each stimulus (click, 6, 12 18, 24, 30 kHz). Plots show the normalized time variance of the averaged signal $S^2(n)/S^2(N)$ (y-axis) vs.  the total percentage of ABR curves included in the cumulative average $n/N$ (x-axis). a) shows NN predictions and SLR estimations from experiments 1 and 5, b) shows NN predictions and SLR estimations from experiments 4 and 8, as introduced in \mbox{table \ref{tab:8experiments}}. Two methods can be compared in a way that the curve of the better method stays longer close to zero.}
\label{fig:eval_curves}
\end{figure}

\mbox{Fig. \ref{fig:eval_curves}} shows evaluation curves for data from experiments 1 and 5 (GMC) and from experiments 4 and 8 (ING). Evaluation curves of cross-over experiments 2, 6, 3 and 7 are shown in \mbox{supplement table \ref{fig:sup_eval_curves}}. All methods begin to deviate from zero quite early, so none of them seems to be perfect. However, curves show that for GMC data, both NN and SLR outperform manual threshold finding, with NN overall being slightly better than SLR. In contrast, for ING data, the three methods (human, NN, SLR) differ only marginally, with SLR overall being best. 

Using evaluation curves as an unbiased tool, it can be concluded that human threshold finding cannot automatically assumed to be the best method. Data sets may exhibit different levels of variability and human bias. In this regard, the ING data set is more consistent than the GMC data set, which only underpins the need for unbiased threshold finding methods.

Results from evaluation curves in part contradict the assumptions behind the accuracy based evaluation which treat the human labeled thresholds as ground truth. Obviously, this is not always the case and seems to depend on the level of variability and human bias represented in a data set. When abandoning the premise that human threshold reading always delivers the ground truth, both methods introduced in this work perform very well.

\subsection{Both NN and SLR methods perform well in an end-to-end phenotyping pipeline}

While it is interesting to know that NN and SLR work well for unrelated single stimuli on single mice, using them for routine hearing assessment in high throughput mouse phenotyping is a different matter. In such a scenario, found thresholds are usually aggregated on two levels: first, thresholds for all stimuli of one individual are aggregated to a hearing curve, then, hearing curves are aggregated to display mutant vs. control threshold medians or means.

To find out whether NN and SLR are able to identify mouse lines with biologically relevant changes in such a scenario, the following approach was applied: complete raw data from both data sets was subjected to NN and SLR threshold finding. However, for downstream gene-based analysis steps, data from some mice had to be excluded. In the GMC data set, 45 mutants without clearly assigned reference controls and in the ING data set, 48 mice without valid gene label were affected.

\begin{figure}[h]
\centering
\includegraphics[width=0.9\linewidth]{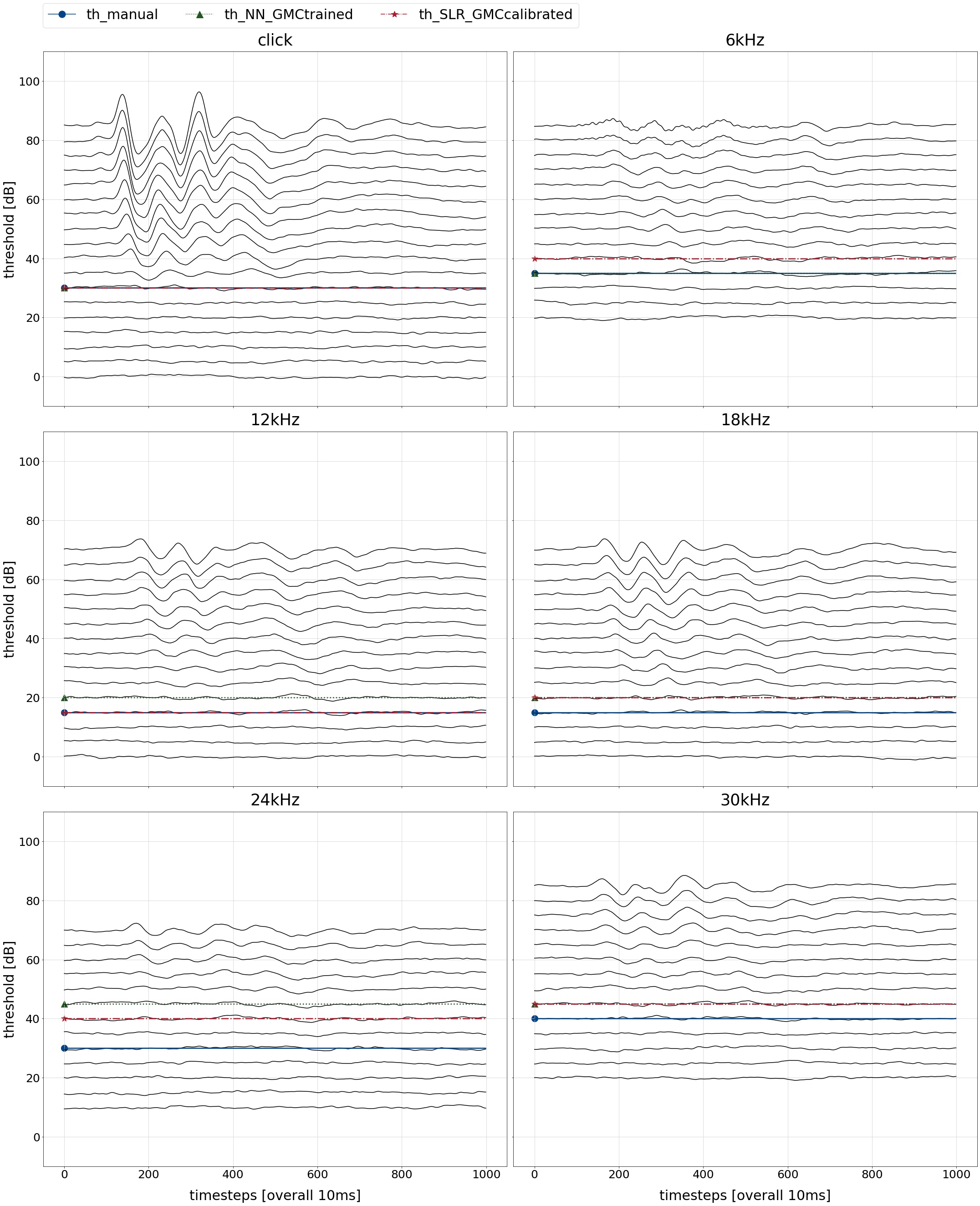}
\caption{\textbf{Audiograms and hearing thresholds of a single GMC mouse}. For all six stimuli, the stacked averaged response signals of an individual mouse are shown. The x-axis covers a time span of \mbox{10 ms} in 1,000 time steps. The y-axis shows stacked response signal strengths, with each curve corresponding to a sound pressure level. Ticks in 20 dB steps indicate where each SPL curve begins. Overlaid horizontal lines indicate hearing thresholds assigned by three methods: manual, by GMC reader (blue, solid line and circle); NN, GMC-trained (green, dotted line and triangle); SLR, GMC-calibrated (red, dashed line and star).}
\label{fig:ABR_signal_single_mouse}
\end{figure}

\paragraph*{Visual identification of candidate genes} \mbox{} \\ Using resulting thresholds, a series of high-level visualisation have been generated that can be used for visual identification of candidate genes. \mbox{Fig. \ref{fig:ABR_signal_single_mouse}} shows an example of an audiogram, which has been generated for every single mouse in the data sets. For all six stimuli, ABR responses as well as respective manual, NN, and SLR thresholds are plotted. 

\begin{figure}[h]
\centering
\includegraphics[width=0.95\linewidth]{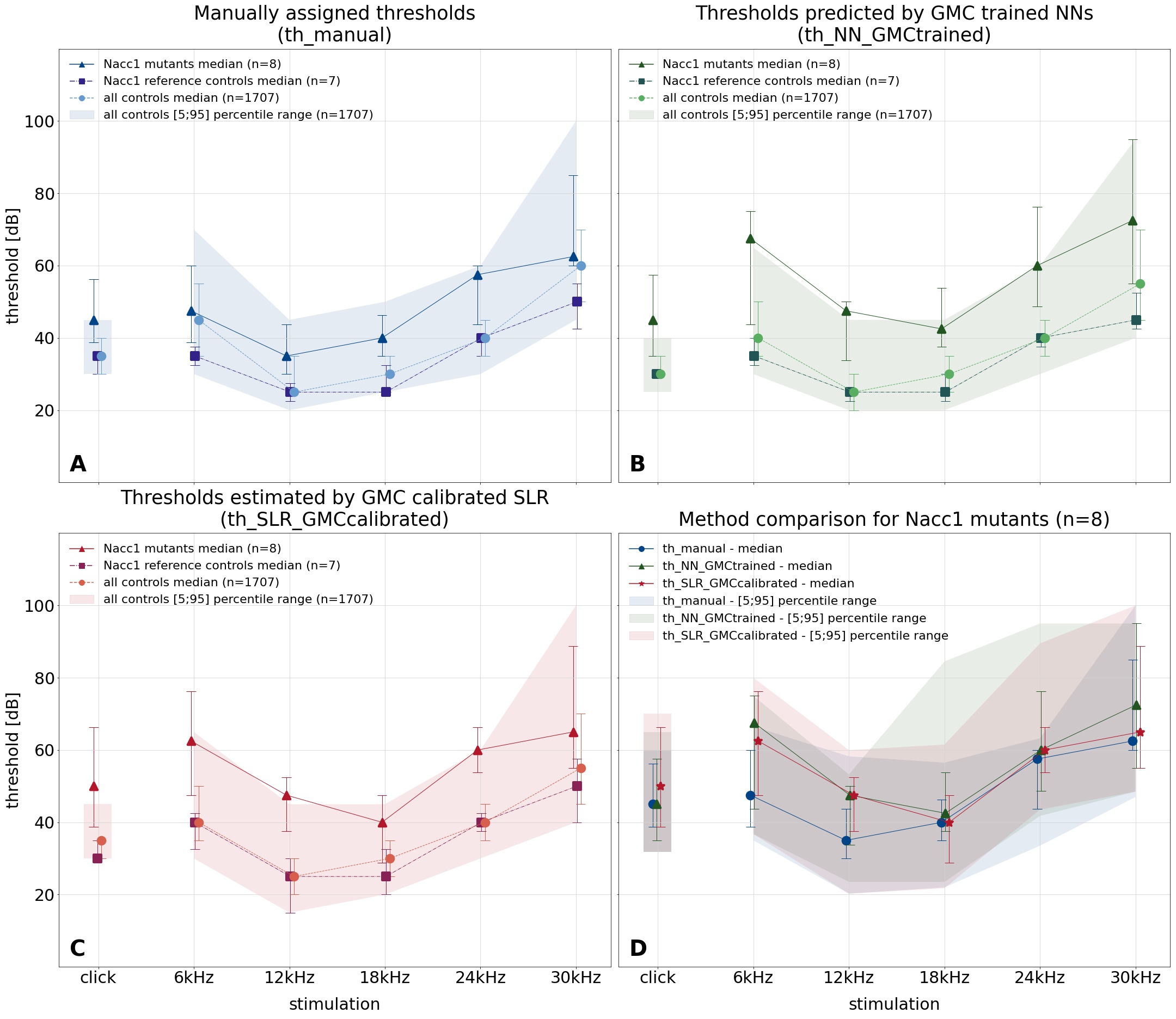}
\caption{\textbf{Group-based comparison of three threshold finding methods}. On the example of the $Nacc1^{em1(IMPC)Hmgu}$ mouse line, the figure shows an end-to-end comparison of the three threshold finding methods for the use case of identifying candidate genes with hearing function involvement. For each method (manual: top left, NN: top right, SLR: bottom left), stimulus-specific hearing thresholds are compiled to median hearing curves of mutants (triangles and solid lines), same-day reference controls (squares and dashed lines) and ``all controls'' (circles and dotted lines) and shown as symbols, which are connected for the non-click stimuli. The ribbon shows the [5;95] percentile range of all controls. Numbers of mice in each group are shown in the legend. On the y-axis, hearing threshold is given in 20 dB ticks. The x-axis shows the stimulus. Whiskers indicate inter-quartile ranges. Each method-specific plot allows comparison of mutant vs. control hearing curves with thresholds based on that method. The bottom-right figure shows mutant-only median hearing curves and [5;95] percentile ranges of the three methods (manual: blue, solid line and circle; NN: green, dotted line and triangle; SLR: red, dashed line and star) to allow method comparison.}
\label{fig:curve_group_comp}
\end{figure}

Next, for all GMC lines, hearing curves were generated that show mutant vs. control group medians, with a background indicating the [5;95] percentile range of all control animals. This is done in separate subplots for manual, NN, and SLR thresholds, to allow comparison of hearing curve differences of mutants and controls between methods. A forth subplot only shows overlaid mutant median hearing curves for all three methods. \mbox{Fig. \ref{fig:curve_group_comp}} shows on the example of the $Nacc1^{em1(IMPC)Hmgu}$ mouse line, that all methods are able to detect the shift of the hearing curve in mutants. This use case shows a clear advantage of the algorithmic methods: there may be a systematic shift with regards to the manual method. However, it applies to both mutants and controls, conserving any differences between both. Both methods can also be considered blinded, as they are not aware to which group an ABR response signal belongs. 

\begin{figure}[h]
\centering
\includegraphics[width=0.9\linewidth]{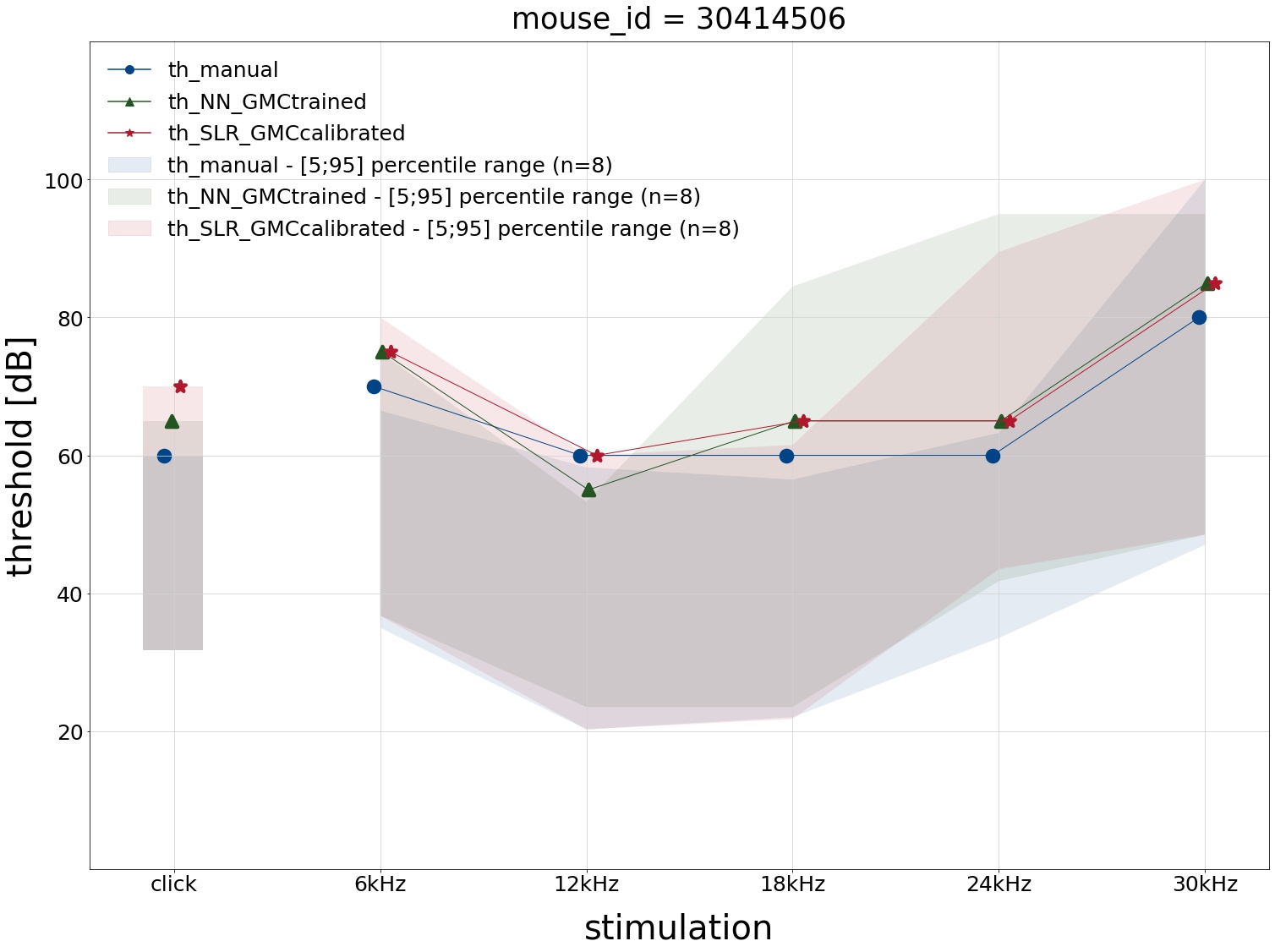}
\caption{\textbf{Mouse-based comparison of three threshold finding methods}. On the example of an individual $Nacc1^{em1(IMPC)Hmgu}$ mutant mouse, the figure shows an end-to-end comparison of the three threshold finding methods. Stimulus-specific hearing thresholds are displayed as hearing curves gained by manual (blue, solid line and circle), NN (green, dotted line and triangle) and SLR (red, dashed line and star) threshold finding. The ribbons show the respective [5;95] percentile range of all mutants of the same gene in the respective colour. On the y-axis, hearing threshold is given in 20 dB ticks. The x-axis shows the stimulus. }
\label{fig:curve_single_mouse}
\end{figure}

Finally, for each mouse, another plot shows an overlay of hearing curves for the three methods in comparison. \mbox{Fig. \ref{fig:curve_single_mouse}} shows an example of a $Nacc1^{em1(IMPC)Hmgu}$ mouse where all three methods agree quite well. 

All plots are made publicly available and can be used to validate and compare the methods on the original data. 

\begin{figure}[h]
\centering
\includegraphics[width=0.95\linewidth]{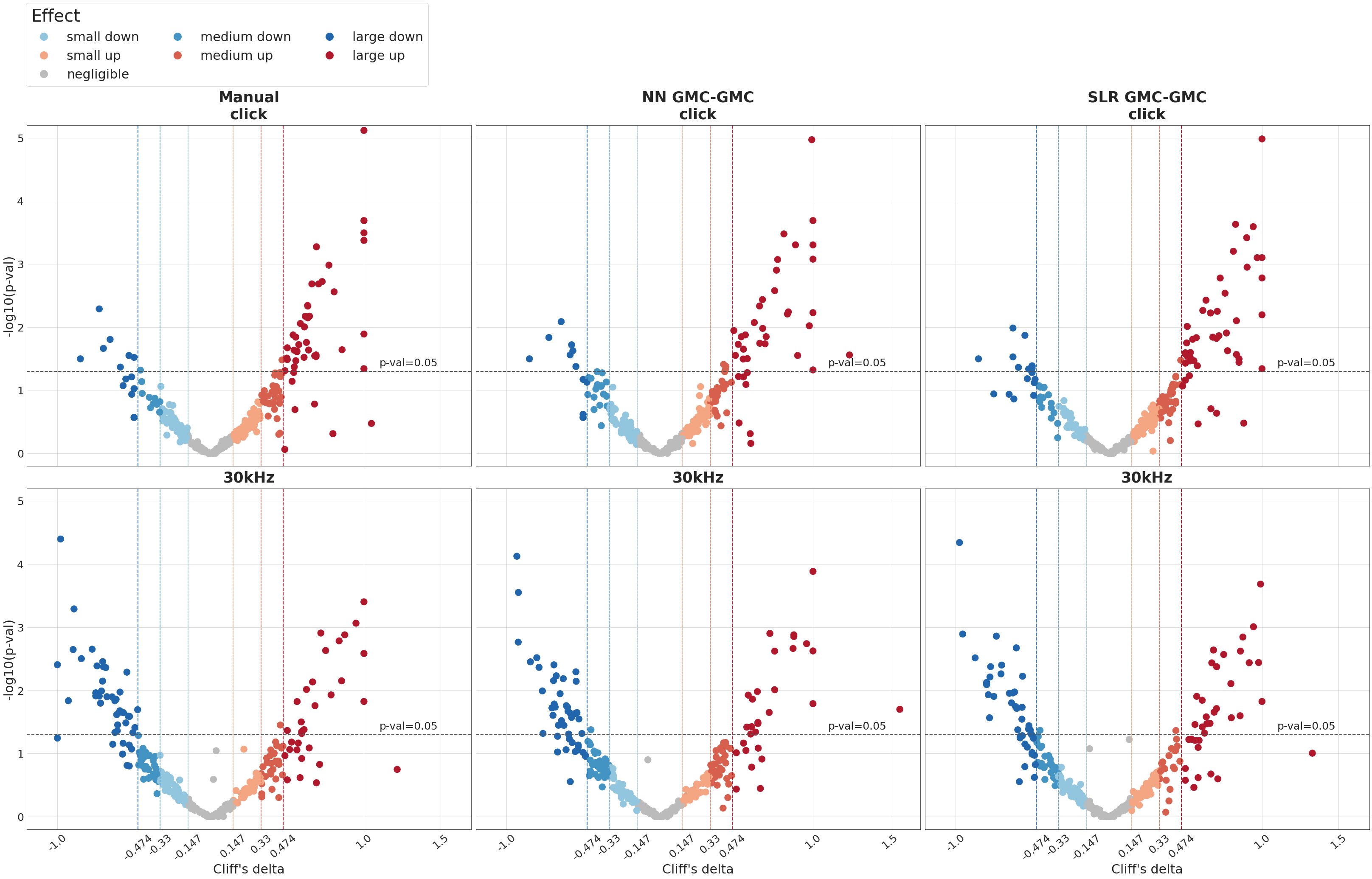}
\caption{\textbf{Biologically relevant changes in hearing thresholds - GMC lines, click and 30 kHz}. Volcano plots show significance vs. effect size for all GMC lines. For each mouse line, represented by a dot, hearing thresholds were used to calculate significance (Wilcoxon rank sum test, y-axis) and non-parametric effect size (Cliff's delta \autocite{cliff1993}, x-axis) of mutant vs. control animals.  Vertical lines indicate margins for small (0.147), medium (0.33) and large (0.474) effects as suggested in \autocite{romano2006}. The horizontal line indicates the 0.05 significance threshold level. Accordingly, dots in the upper left and upper right areas denote GMC lines with significant as well as relevant changes and thus are considered worthwhile candidates (see supplement tables \ref{tab:top_candidates_click} and \ref{tab:top_candidates_30kHz}). Dot colours in addition represent effect size as shown in the legend. Plot rows represent click (upper) and 30 kHz (lower) stimulus data. Columns compare the three hearing threshold finding methods compared in this work (left: manual, middle: NN, right: SLR).}
\label{fig:GMC_volcano_plot_click_30kHz}
\end{figure}

\paragraph*{Fully automated identification of candidate genes} \mbox{} \\ Visual comparison of hearing curves is indispensable for evaluation purposes, however not feasible for screening, since it is laborious and, similar to curve reading, it may be prone to bias. Therefore, a programmatic approach has been implemented that uses two measures as criteria to detect mutant mouse lines that exhibit potential biologically meaningful changes in hearing. First, effect size, which descriptively spoken measures the degree of overlap between mutant and control group distribution of a stimulus-specific threshold. As no normal distribution can be assumed, Cliff's Delta was used, which ranges between \mbox{-1 and 1}. Second, significance, using p-values resulting from a Wilcoxon rank sum test, defined as the probability of getting a test statistics as large or larger assuming mutant and control distribution are the same. A well-established way of displaying these two measures is the so-called volcano plot. \mbox{Fig. \ref{fig:GMC_volcano_plot_click_30kHz}} shows such volcano plots for click and 30 kHz thresholds of GMC lines for all three methods. Here, interesting lines - i.e. lines that exhibit a biologically meaningful hearing phenotype - are supposed to be those that show high significance and a large effect size at the same time. Using $p<0.05$ and $|d|>0.474$ for large effects \autocite{romano2006}, candidate mouse lines can be found in the upper left (lower threshold) and upper right (higher threshold) area of the plots and of course can be directly filtered to result lists.

\begin{figure}[h]
\centering
\includegraphics[width=0.95\linewidth]{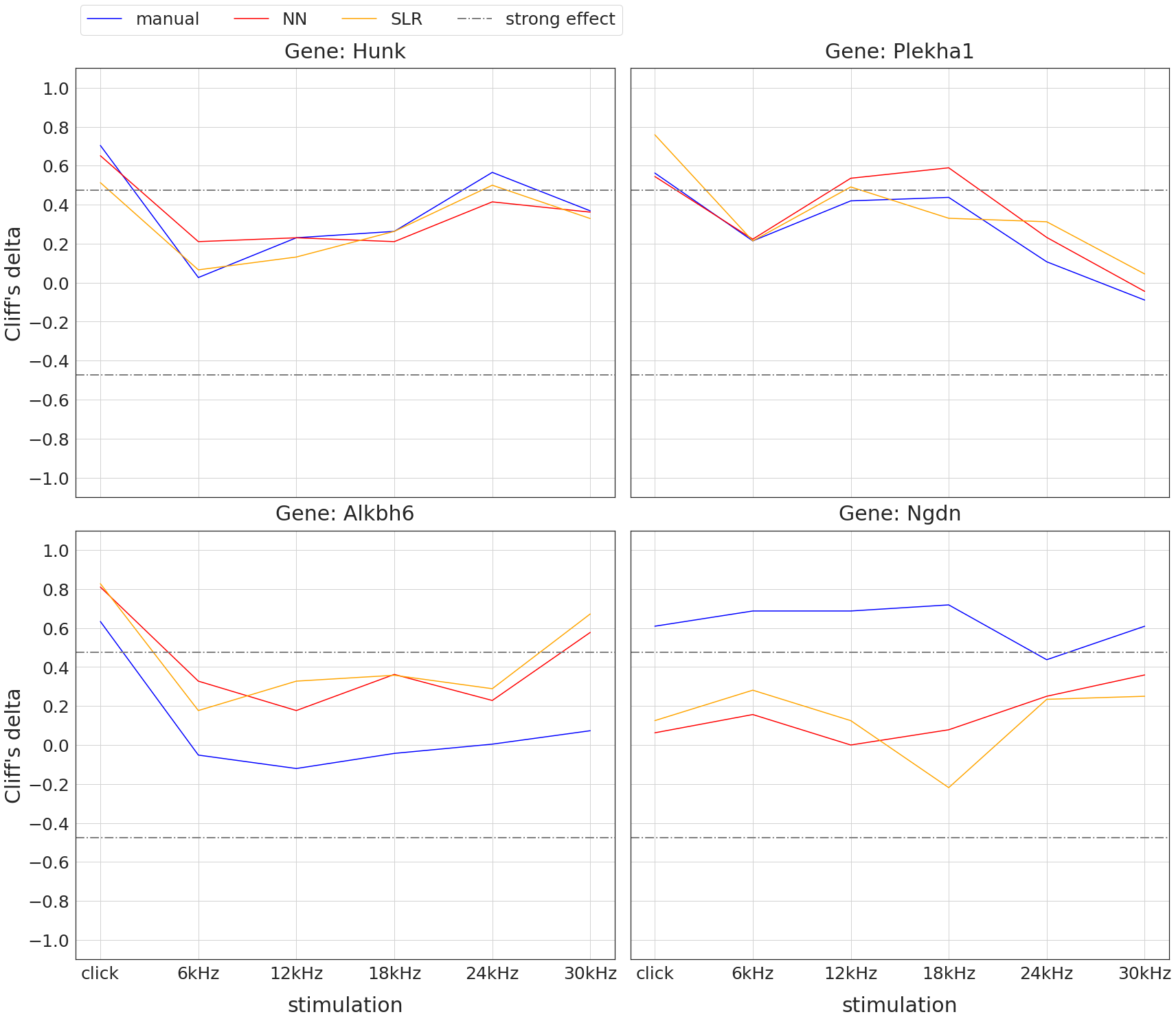}
\caption{\textbf{Comparison of mutant vs. control effect sizes between three hearing threshold finding methods}. For each of four selected genes with strong effects, plots show non-parametric effect size (Cliff's delta \autocite{cliff1993}, y-axis) of mutant vs. control animals for all six stimuli (x-axis). Colours indicate the threshold finding method (manual: blue, NN: red, SLR: orange). The grey dashdotted horizontal line shows the threshold for strong effects at 0.474. For convenience, lines connect effect sizes of the same method (Note: this is not a hearing curve, click has to be interpreted separately). For the two genes in the top row (\textit{Hunk, Plekh2}), effect sizes differ only marginally for all stimuli. Bottom left shows an example (\textit{Alkbh6}) where larger effects are found with NN and SLR throughout all stimuli. In contrast, bottom right shows an example (\textit{Ngdn}) where smaller effects are found with NN and SLR throughout all stimuli.}
\label{fig:strong_effect_4genes}
\end{figure}

Supplementary tables \ref{tab:top_candidates_click} and \ref{tab:top_candidates_30kHz} each show a method comparison of top candidate lines/genes for modified threshold at click and 30 kHz stimulus, respectively. Not surprisingly, lists are largely similar, although not completely. For example, all three methods identified \textit{Gpsm2} as well as \textit{Rest}, two well-known hearing loss genes \autocite{hereditaryhearingloss}, while other hits differ at least at single frequencies. To further improve facilitated identification of candidate genes, a new plot displays calculated effect sizes for all stimuli and all three methods. \mbox{Fig. \ref{fig:strong_effect_4genes}} shows on four examples of this highly integrated plot, that it allows to rapidly evaluate ABR results in two ways: a) assess effect sizes for the different stimuli and thus judge the nature of hearing impairment, b) compare effect sizes derived from different methods. As can be seen for genes \textit{Hunk} and \textit{Plekha1}, all three methods end up in almost identical effect sizes and overall pattern. For two other lines, automated methods differ from human threshold finding in delivering consistently larger (\textit{Alkbh6}) and smaller (\textit{Ngdn}) effects.

\begin{figure}[h]
\centering
\includegraphics[width=0.95\linewidth]{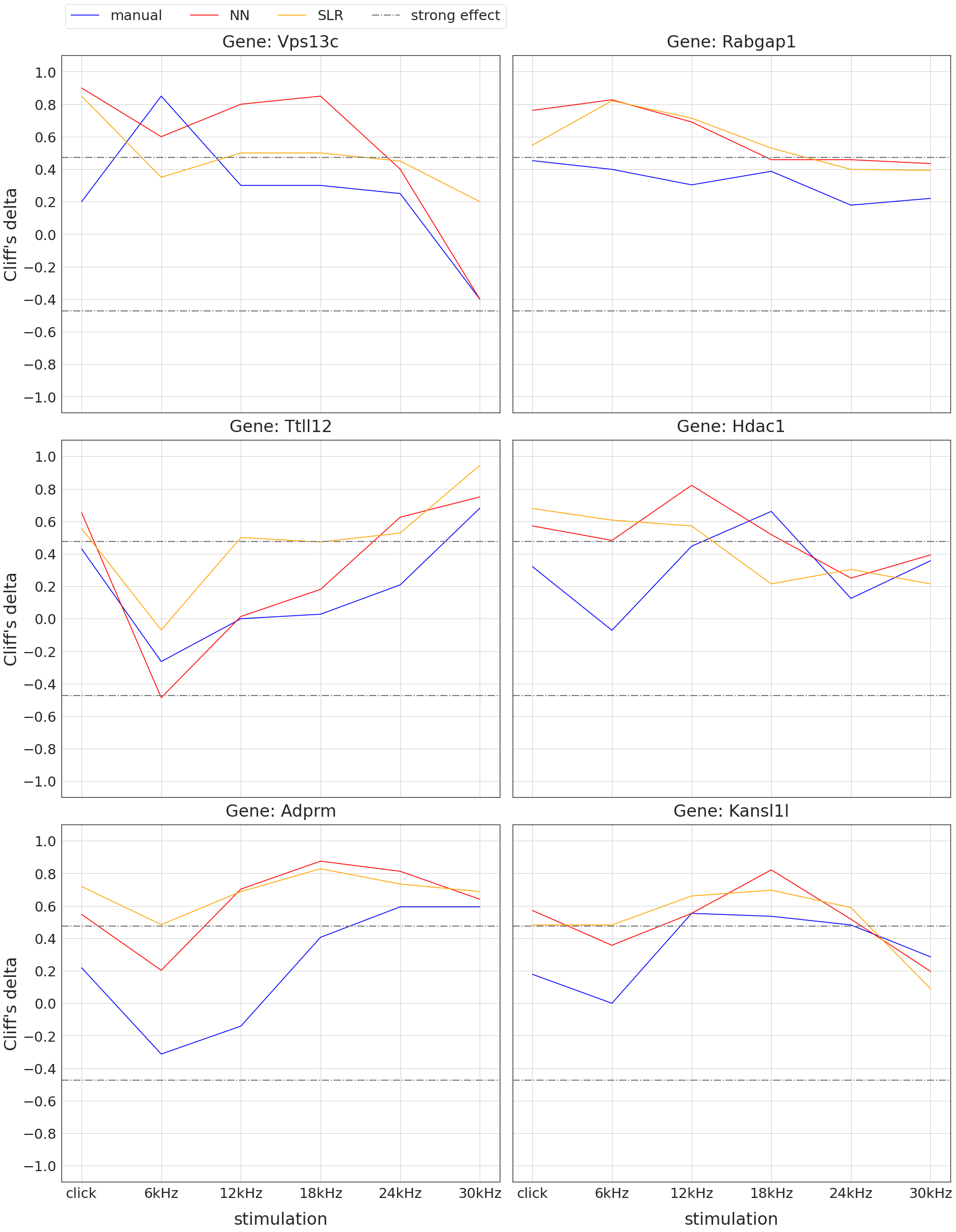}
\caption{\textbf{Visual identification of six new candidate genes with hearing impact at click stimulation}. Plots show non-parametric effect size (Cliff's delta \autocite{cliff1993}, y-axis) of mutant vs. control animals for all six stimuli (x-axis). Colours indicate the threshold finding method (manual: blue, NN: red, SLR: orange). The grey dashdotted horizontal lines show the thresholds for strong effects at $\pm$0.474. For convenience, lines connect effect sizes of the same method (Note: this is not a hearing curve, click has to be interpreted separately). For each of the six genes shown, click effect sizes of NN and SLR derived thresholds are above the dashdotted line, indicating strong effects, whereas manual threshold effects are below.}
\label{fig:strong_effect_6genes}
\end{figure}

\begin{figure}[h]
\centering
\includegraphics[width=0.95\linewidth]{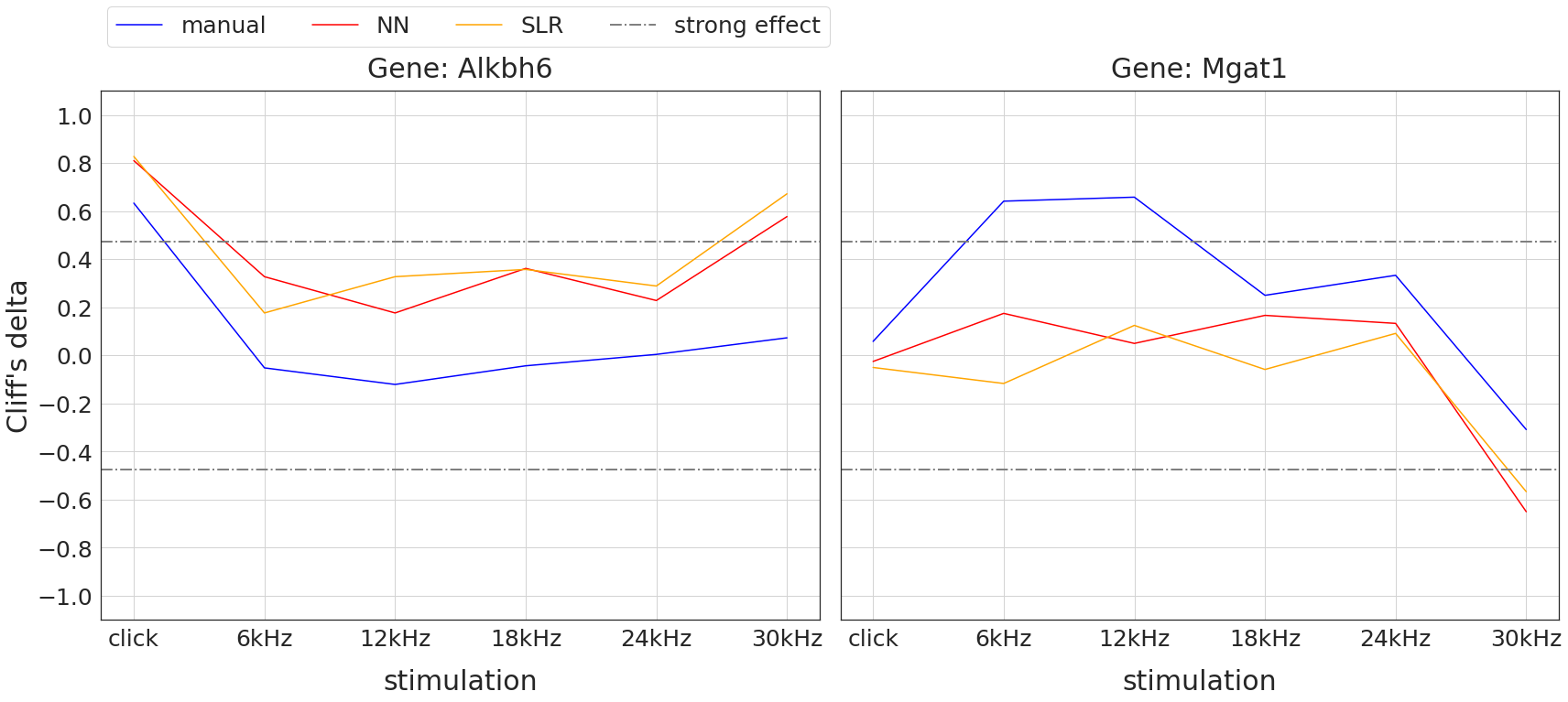}
\caption{\textbf{Visual identification of two new candidate genes with hearing impact at 30 kHz stimulation}. Plots show non-parametric effect size (Cliff's delta \autocite{cliff1993}, y-axis) of mutant vs. control animals for all six stimuli (x-axis). Colours indicate the threshold finding method (manual: blue, NN: red, SLR: orange). The grey dashdotted horizontal lines show the thresholds for strong effects at $\pm$0.474. For convenience, lines connect effect sizes of the same method (Note: this is not a hearing curve, click has to be interpreted separately). For the two genes shown, 30 kHz effect sizes of NN and SLR derived thresholds are above the $+0.474$ (left: \textit{Alkbh6}) and below the $-0.474$ (right: \textit{Mgat1}) dashdotted line, indicating strong effects, whereas manual threshold effects don't.}
\label{fig:strong_effect_2genes}
\end{figure}

\begin{figure}[h]
\centering
\includegraphics[width=0.95\linewidth]{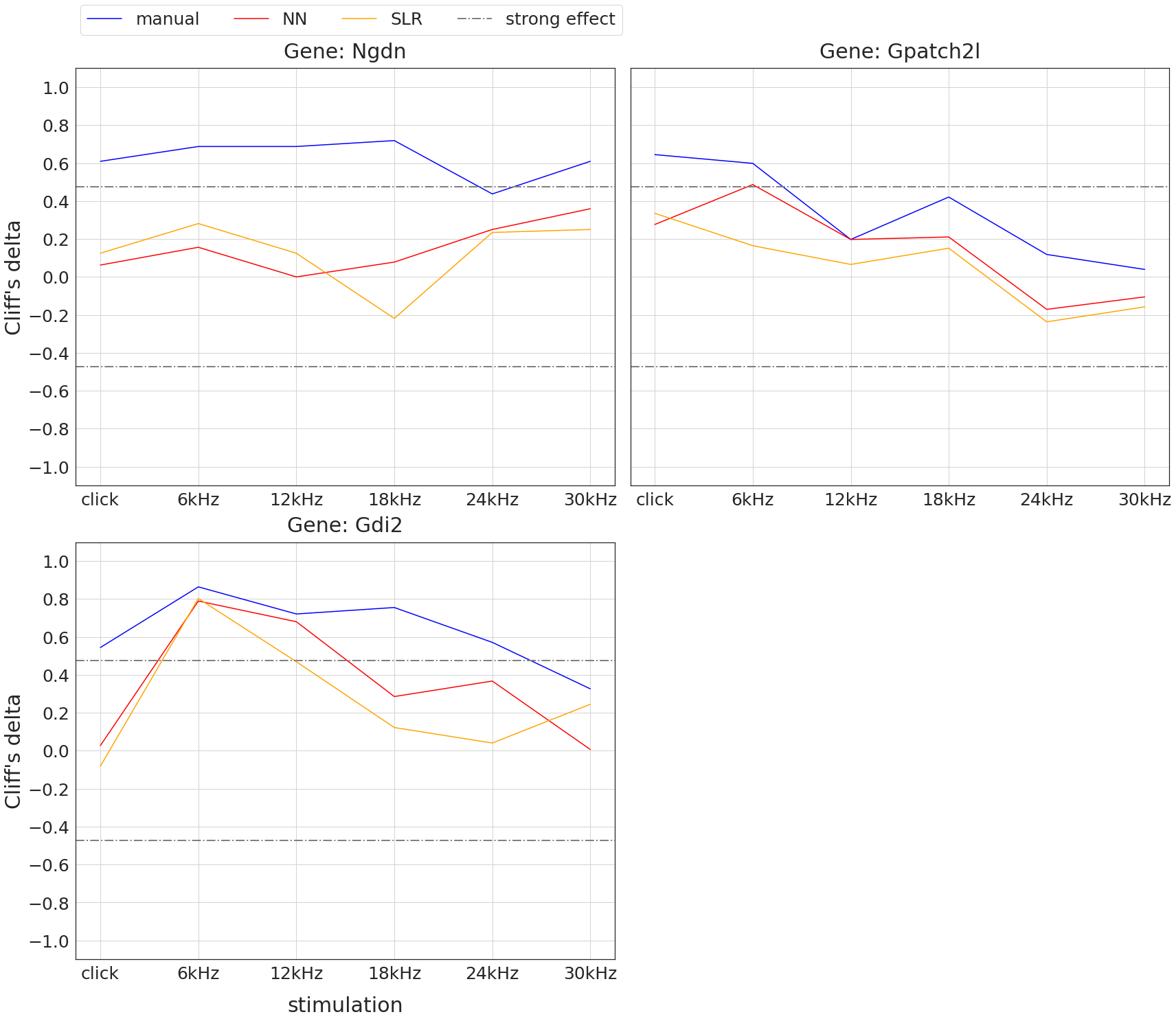}
\caption{\textbf{Loss of hearing sensitivity gene candidate status using automated threshold finding}. Plots show non-parametric effect size (Cliff's delta \autocite{cliff1993}, y-axis) of mutant vs. control animals for all six stimuli (x-axis). Colours indicate the threshold finding method (manual: blue, NN: red, SLR: orange). The grey dashdotted horizontal lines show the thresholds for strong effects at $\pm$0.474. For convenience, lines connect effect sizes of the same method (Note: this is not a hearing curve, click has to be interpreted separately). For the three genes shown, click and/or 30 kHz manual threshold effect sizes are above the $+0.474$ dashdotted line, indicating strong effects, whereas SLR and NN threshold effects don't.}
\label{fig:strong_effect_3genes}
\end{figure}

\paragraph*{An end-to-end analysis pipeline using SLR based thresholds reveals 76 candidate genes with impact on hearing sensitivity} \mbox{} \\ In a re-analysis of the GMC raw data set, hearing thresholds derived from both automated methods (SLR and NN) were used for identification of candidate lines as described above. For click stimulation, the visual and/or the fully automated method identified six genes (\textit{Vps13c, Rabgap1, Ttll12, Hdac1, Adprm, and Kansl1l}, Fig. \ref{fig:strong_effect_6genes}) with strong effects that had not been detected so far using manual thresholds only. For 30 kHz stimulation, two new candidate genes (\textit{Alkbh6 and Mgat1}, Fig. \ref{fig:strong_effect_2genes}) were identified. For a set of three other manually identified candidate genes (\textit{Ngdn, Gpatch2l, Gdi2}, Fig. \ref{fig:strong_effect_3genes}), NN and SLR derived thresholds did not lead to strong effects at click or 30 kHz.

Of course, evaluation of hearing deficits is not relying on differences at single frequencies. For identifying a number genes with impact on hearing sensitivity, the evaluation of single thresholds is the basis for analysis. Additional steps will include the definition of relevant effect sizes and patterns of alteration. To further explore these potential hearing genes, databases for human variants, expression patterns, pathways etc. will have to strengthen the evidence for candidate genes. In addition, confirmation of results with calculated sample sizes and/or separation of sexes is needed in some cases.

Altogether, 76 potential hearing genes have been detected by automated analysis starting from raw data using SLR (see supplemental tables \ref{tab:top_candidates_click} and \ref{tab:top_candidates_30kHz}, unique entries from combined SLR columns). For four of them (\textit{Hoxa2, Aspa, Gpsm2}, and \textit{Rest}), human orthologue genes have published annotations for human hearing loss  according to {OMIM}{\textsuperscript{\textregistered}} \autocite{omim}. Inner ear gene expression  was evaluated by literature \autocite{scheffer2015, ranum2019} and eleven of the genes were reported to be expressed in hair cells or surrounding cells. For 35 of the genes, no mouse model was yet listed at the Mouse Genome Database (MGD) \autocite{mgd2019}, while for 37 of them with a mouse model available no information about hearing sensitivity was provided. Solely for four of the mouse models, either altered hearing or middle ear morphology was reported (\textit{Rest, Gpsm2, Aspa}, and \textit{Hoxa2}). Some of the genes are already associated with human disease, underlining the pleiotrophy of gene functions and phenotypes. For example, \textit{Btbd 9} is associated with restleg legs syndrome (RLS, OMIM 611185), but is also expressed in outer hair cells \autocite{ranum2019}, thus providing a possible link to the detected hearing alteration. Further analysis will be needed for the possible candidate genes to uncover the nature of gene-phenotype association.

\section{Conclusions}

Using two independent and large data sets, this work shows that two new methods are robust and able to objectively detect hearing thresholds from averaged ABR raw data. While the supervised NN method, using two neural networks, achieves higher accuracies for manual ground truth, it requires training with large numbers of human-assigned labels and cannot be transferred between data sets. Thus, it may be preferred by large laboratories with high level manual thresholding standards. The self-supervised Sound Level Regression - SLR - method does not depend on labels and thus can be directly applied to any ABR data set. 

Both methods have the advantage of delivering highly consistent results. As they can be employed in fully-integrated end-to-end pipelines, they are predestined for use in routine measurements, quality control, and automated retrospective re-analysis of large ABR data collections. 
Since SLR is invariant to the data set, it offers itself as a method for meta analysis of ABR data from different institutions. 

In a mutant screening environment, both NN and SLR can be integrated into a fully automated end-to-end pipeline, starting from raw averaged ABR data and finally producing candidate lists and plots. 

The decision to trust NN- and SLR-derived thresholds over manual derived thresholds is subjective. However, this work - using two independent data sets - supplies a solid foundation of data, results and comparative plots for everyone to allow an informed decision. In addition, the provided methods allow comparative analysis of all methods using own data. 

Code of this work is available at \url{https://github.com/ExperimentalGenetics/ABR\_thresholder}. Original raw and intermediate data, results, and all generated plots are available at zenodo.org for review and use (\url{http://dx.doi.org/doi:10.5281/zenodo.5779876}).

\section*{Declaration of interest}
The authors of this manuscript have no potential financial, personal or other conflicts with other people or organisations that could inappropriately influence their work.

\section*{Author contributions}
\textbf{DT:} Conceptualisation, Methodology, Software, Formal analysis, Writing - Original Draft, Visualisation 
\textbf{ES:} Software, Validation, Data Curation, Visualisation
\textbf{GM:} Conceptualisation, Methodology, Software, Formal analysis, Data Curation
\textbf{AH:} Software, Visualisation
\textbf{MHdA:} Resources, Supervision, Funding acquisition, Writing - Review \& Editing	
\textbf{LB:} Validation, Investigation, Data Curation, Writing - Review \& Editing
\textbf{CLM:} Conceptualisation, Writing - Review \& Editing, Supervision
\textbf{HM:} Conceptualisation, Writing - Original Draft, Writing - Review \& Editing, Supervision, Project administration

\section*{Acknowledgements}

We thank D. Feeser, A. Badmann, R. Fischer, E. Köfferlein, and F. Schleicher for ABR measurements and identification of hearing thresholds. We also thank R. Steinkamp for data capture as well as V. Gailus-Durner and H. Fuchs for critically reading the manuscript. D. Thalmeier and C.L. Müller were funded by Helmholtz Association’s Initiative and Networking Fund through Helmholtz AI.

\section*{Keywords}
automation, auditory brainstem response, evoked potentials, high-throughput hearing screening, objective hearing threshold detection

\printbibliography

\newcommand{\beginsupplement}{%
        \setcounter{table}{0}
        \renewcommand{\thetable}{S\arabic{table}}%
        \setcounter{figure}{0}
        \renewcommand{\thefigure}{S\arabic{figure}}%
     }
\beginsupplement

\clearpage
\section*{Supplement 1 - Neural network model architectures}\label{sup1:model_architectures}
\input{supplements/supplement1}

\clearpage

\section*{Supplement 2 - Evaluation curves}\label{sup2:evaluation_curves}
\input{supplements/supplement2}

\clearpage

\section*{Supplement 3 - Volcano plots of GMC mutant lines}\label{sup3:volcano_plots}
\input{supplements/supplement3}

\clearpage

\section*{Supplement 4 - Comparison of top click candidate genes with hearing threshold changes}\label{sup4:top_candidates_click}
\input{supplements/supplement4}

\clearpage

\section*{Supplement 5 - Comparison of top 30 kHz candidate genes with hearing threshold changes}\label{sup5:top_candidates_30kHz}
\input{supplements/supplement5}

\clearpage

\section*{Supplement 6 - Information on 76 SLR-based candidate genes}\label{sup6:details_76_genes}
\input{supplements/supplement6}

\end{document}

%% file: tables/twoDatasets.tex
\begin{table}[]
\centering
\caption{\textbf{Basic dataset properties.} \textit{A) number of mice:} shown are numbers of distinct, individual mice. In the ING data set, no distinction between male and female numbers was possible, so only total numbers are given. \textit{B) gene cohort size median:} the median size of cohorts of animals with the same affected distinct knockout gene are given along with the 5\% and 95\% quantiles. \textit{C) number of genes:} the number of distinct knockout genes per data set is given. Common genes provides the number of genes occuring in both datasets: \textit{Bach2, Cdkal1, Dbn1, Dnase1l2, Entpd1, Gsk3a, Hdac1, Klk5, Nxn, Rnf10, Slc20a2, Ubash3a}.}
\label{tab:twoDatasets}

\begin{tabular}{r|ccc|ccc}
\multicolumn{1}{c}{}                  & \multicolumn{3}{c|}{\textbf{GMC dataset}}       & \multicolumn{3}{c}{\textbf{ING data set}}               \\ 
\multicolumn{1}{l|}{\textbf{\textit{A) number of mice}}}  & \textbf{mutants} 
                                                          & \textbf{controls} 
                                                          & \textbf{total} 
                                                          & \textbf{mutants} 
                                                          & \textbf{controls} 
                                                          & \multicolumn{1}{c}{\textbf{total}}                                                    \\       
\hline 
\multicolumn{1}{r|}{males}                      & 1,331   & 849      & 2,180            & -       & -        & \multicolumn{1}{c}{-}              \\
\multicolumn{1}{r|}{females}                    & 1,323   & 858      & 2,181            & -       & -        & \multicolumn{1}{c}{-}              \\ 
\hline 
\multicolumn{1}{r|}{total}                      & 2,654   & 1,707    & \textbf{4,361}   & 6,130   & 1,900    &\multicolumn{1}{c}{\textbf{8,030}}  \\ 
\multicolumn{7}{c}{} 
\\

\multicolumn{7}{l}{\textbf{\textit{B) gene cohort size median [5\%;95\%]}}}                                                                       \\    
\hline
\multicolumn{1}{r|}{gene cohort size} & \multicolumn{3}{c|}{8 [3;11]}                   & \multicolumn{3}{c}{4 [4;10]}                            \\ 
\multicolumn{7}{c}{} 
\\

\multicolumn{2}{l}{\textbf{\textit{C) number of genes}}}                                & \multicolumn{5}{l}{}                                    \\
\hline
\multicolumn{1}{r|}{distinct genes}   & \multicolumn{3}{c|}{352}                        & \multicolumn{3}{c}{1,152}                               \\ \cline{2-7}
\multicolumn{1}{r|}{common genes}     & \multicolumn{6}{c}{12}                                                                                    \\ 
\end{tabular}
\end{table}

%% file: tables/8experiments.tex
\definecolor{Lightgray}{gray}{0.95}

\begin{table}[]
\centering
\caption{\textbf{Experiment overview}. Two different ABR treshold finding methods were tested on two different data sets (GMC and ING). The first two columns contain experiments with the two-stage neural network (NN), the last two columns contain experiments with the sound level regression method (SLR). Sub-columns specify the data set that was used for training (NN) or calibration (SLR), respectively. The two rows indicate the data set that was used for testing of the trained NN or the calibrated SLR model. Cells provide the experiment number and the name of the experiment as used in the text. Experiments that use data from the same data set for training/calibration and testing are highlighted in grey. }
\label{tab:8experiments}
\begin{small}
\begin{tabular}{r|cc|cc}
                              & \multicolumn{2}{c|}{\textbf{NN trained on}}                              
                              & \multicolumn{2}{c}{\textbf{SLR calibrated on}}                            \\
\textbf{tested on}            & \textbf{GMC}                        & \textbf{ING}                        
                              & \textbf{GMC}                        & \textbf{ING}                        \\ \hline
\multirow{2}{*}{\textbf{GMC}} & \cellcolor{Lightgray}experiment 1   & experiment 3                        
                              & \cellcolor{Lightgray}experiment 5   & experiment 7                        \\
                              & \cellcolor{Lightgray}``NN GMC-GMC''   & ``NN ING-GMC''                        
                              & \cellcolor{Lightgray}``SLR GMC-GMC''  & ``SLR ING-GMC''                       \\ 
\multirow{2}{*}{\textbf{ING}} & experiment 2                        & \cellcolor{Lightgray}experiment 4\  
                              & experiment 6                        & \cellcolor{Lightgray}experiment 8   \\
                              & ``NN GMC-ING''                        & \cellcolor{Lightgray}``NN ING-ING''   
                              & ``SLR GMC-ING''                       & \cellcolor{Lightgray}``SLR ING-ING''  \\ 
\end{tabular}
\end{small}
\end{table}

%% file: tables/nn_accuracies.tex
\begin{table}[]
\centering
\caption{\textbf{NN accuracies for two data sets, stimuli and match levels}. Major columns \mbox{``NN GMC-GMC''} and \mbox{``NN ING-ING''} correspond to two experiments introduced in table \ref{tab:8experiments}. The ``exact'' columns contain accuracy values when requiring exact match of human-assigned threshold label and NN prediction. The \mbox{``$\pm$5 dB''} and \mbox{``$\pm$10 dB''} columns contain accuracy values when allowing 5 dB and 10 dB tolerance, respectively. Numbers in cells denote the accuracy of the model prediction at the stimulus and match level.}
\label{tab:nn_accuracies}
\begin{small}
\begin{tabular}{r|rrr|rrr}
\multicolumn{1}{c}{}  & \multicolumn{3}{c}{\textbf{NN GMC-GMC}}                    & \multicolumn{3}{c}{\textbf{NN ING-ING}}                    \\ 
\multicolumn{1}{c}{}  & \multicolumn{3}{c}{\textbf{accuracy [\%]}}                 & \multicolumn{3}{c}{\textbf{accuracy [\%]}}                 \\
\textbf{stimulus}     & \textbf{exact} & \textbf{$\pm$5 dB} & \textbf{$\pm$10 dB}  & \textbf{exact} & \textbf{$\pm$5 dB} & \textbf{$\pm$10 dB}  \\ \hline 
\textbf{click}        & 19.5           & 90.3               & 98.5                 & 12.6           & 83.9               & 99.3                 \\ 
\textbf{6 kHz}        & 28.8           & 70.3               & 87.9                 & 22.0           & 77.1               & 94.9                 \\ 
\textbf{12 kHz}       & 32.8           & 80.1               & 93.9                 & 22.2           & 79.5               & 95.9                 \\ 
\textbf{18 kHz}       & 28.3           & 78.4               & 93.4                 & 15.1           & 75.3               & 96.0                 \\ 
\textbf{24 kHz}       & 25.0           & 73.9               & 88.0                 & 14.1           & 76.9               & 96.6                 \\ 
\textbf{30 kHz}       & 21.6           & 60.5               & 77.4                 & 16.5           & 73.6               & 95.4                 \\  \hline
\textbf{overall}      & \textbf{26.0}  & \textbf{75.6}      & \textbf{89.8}        & \textbf{17.1}  & \textbf{77.7}      & \textbf{96.3}                         
\end{tabular}
\end{small}
\end{table}

%% file: tables/slr_accuracies.tex
\begin{table}[]
\centering
\caption{\textbf{SLR accuracies for two data sets, stimuli and match levels}. Major columns \mbox{``SLR GMC-GMC''} and \mbox{``SLR ING-ING''} correspond to two experiments introduced in table \ref{tab:8experiments}. The ``exact'' columns contain accuracy values when requiring exact match of  human-assigned threshold label and SLR estimation. The \mbox{``$\pm$5 dB''} and \mbox{``$\pm$10 dB''} columns contain accuracy values when allowing 5 dB and 10 dB tolerance, respectively. Numbers in cells denote the accuracy of the model estimation at the stimulus and match level.}
\label{tab:slr_accuracies}
\begin{small}
\begin{tabular}{r|rrr|rrr}
\multicolumn{1}{c}{}  & \multicolumn{3}{c}{\textbf{SLR GMC-GMC}}                     & \multicolumn{3}{c}{\textbf{SLR ING-ING}}                      \\ 
\multicolumn{1}{c}{}  & \multicolumn{3}{c}{\textbf{accuracy [\%]}}                   & \multicolumn{3}{c}{\textbf{accuracy [\%]}}                    \\
\textbf{stimulus}     & \textbf{exact} & \textbf{$\pm$5 dB}   & \textbf{$\pm$10 dB}  & \textbf{exact}    & \textbf{$\pm$5 dB} & \textbf{$\pm$10 dB}  \\ \hline 
\textbf{click}        & 59.5          & 95.4            & 98.7                & 44.0           & 91.2            & 98.3              \\ 
\textbf{6 kHz}        & 24.4          & 58.7            & 79.8                & 14.3           & 48.9           & 74.5              \\ 
\textbf{12 kHz}       & 27.7          & 66.4            & 85.9                & 17.5           & 54.1            & 79.6              \\ 
\textbf{18 kHz}       & 30.4          & 69.9            & 89.7                & 18.7           & 58.2            & 83.6              \\ 
\textbf{24 kHz}       & 33.5          & 73.3            & 89.3                & 18.7           & 58.5            & 84.2              \\ 
\textbf{30 kHz}       & 35.7          & 69.0            & 84.7                & 19.2           & 58.4            & 83.2              \\ \hline
\textbf{overall}      & \textbf{35.2} & \textbf{72.1}   & \textbf{88.0}     & \textbf{22.0}  & \textbf{61.5}   & \textbf{83.9}                      
\end{tabular}
\end{small}
\end{table}

%% file: tables/8experiments_accuraries.tex
\begin{table}[]
\centering
\thisfloatpagestyle{empty}  
\caption{\textbf{Accuracy overview}. For eight experiments, as introduced in \mbox{table \ref{tab:8experiments}}, the table shows overall and stimulus-specific prediction accuracies. In short, columns determine the applied method (NN or SLR) and the training/calibration data set. The header columns denote the data set that was used for testing and the stimulus, respectively. Cells contain the accuracy values. To facilitate interpretation, the best accuracy in each row is marked in bold. Three blocks correspond to the required match level for accuracy calculation: A) exact match, B) $\pm5$ dB, and C) $\pm10$ dB tolerance.}
\label{tab:8experiments_accuraries}
\begin{small}
\begin{tabular}{rr|rr|rr}

\multicolumn{2}{c}{}              & \multicolumn{2}{c}{\textbf{Neural Network (NN)}} 
                                  & \multicolumn{2}{c}{\textbf{Sound level regression (SLR)}}   \\ 
\multicolumn{1}{c}{\textbf{test}} & \multicolumn{1}{c}{}              
                                  & \multicolumn{2}{c}{\textbf{trained on}} 
                                  & \multicolumn{2}{c}{\textbf{calibrated on}}            \\
\multicolumn{1}{c}{\textbf{data}} & \multicolumn{1}{c}{\textbf{stimulus}} 
                                  & \multicolumn{1}{c}{\textbf{GMC}} 
                                  & \multicolumn{1}{c}{\textbf{ING}}  
                                  &  \multicolumn{1}{c}{\textbf{GMC}} 
                                  & \multicolumn{1}{c}{\textbf{ING}}                      \\ \hline \\
\multicolumn{2}{l|}{\cellcolor{Lightgray}\textbf{\textit{A) exact match}}}   
                                  & \multicolumn{1}{c}{\textbf{experiment 1}} 
                                  & \multicolumn{1}{c|}{\textbf{experiment 3}} 
                                  & \multicolumn{1}{c}{\textbf{experiment 5}} 
                                  & \multicolumn{1}{c}{\textbf{experiment 7}}             \\ \hline
\multirow{7}{*}{\textbf{\begin{tabular}[r]{@{}r@{}}GMC\\data\end{tabular}}} 
       & \multicolumn{1}{r|}{\textbf{Overall}} & 26.0 \%          & 9.7 \%   & 35.2 \%          & \textbf{36.0 \%}   \\ 
       & \multicolumn{1}{r|}{\textbf{Click}}   & 19.5 \%          & 16.1 \%  & \textbf{59.5 \%} & 58.5 \%            \\ 
       & \multicolumn{1}{r|}{\textbf{6 kHz}}   & \textbf{28.8 \%} & 10.7 \%  & 24.4 \%          & 24.1 \%            \\ 
       & \multicolumn{1}{r|}{\textbf{12 kHz}}  & \textbf{32.8 \%} & 5.5 \%   & 27.7 \%          & 29.7 \%            \\ 
       & \multicolumn{1}{r|}{\textbf{18 kHz}}  & 28.3 \%          & 7.0 \%   & 30.4 \%          & \textbf{32.7 \%}   \\  
       & \multicolumn{1}{r|}{\textbf{24 kHz}}  & 25.0 \%          & 9.8 \%   & \textbf{33.5 \%} & \textbf{33.5 \%}   \\ 
       & \multicolumn{1}{r|}{\textbf{30 kHz}}  & 21.6 \%          & 9.1 \%   & 35.7 \%          & \textbf{37.1 \%}   \\ \cline{3-6}
\multicolumn{2}{l|}{}             & \multicolumn{1}{c}{\textbf{experiment 2}} 
                                  & \multicolumn{1}{c|}{\textbf{experiment 4}} 
                                  & \multicolumn{1}{c}{\textbf{experiment 6}} 
                                  & \multicolumn{1}{c}{\textbf{experiment 8}}                                        \\ \cline{3-6}
\multirow{7}{*}{\textbf{\begin{tabular}[r]{@{}r@{}}ING\\data\end{tabular}}} 
       & \textbf{Overall}                      & 17.6 \%           & 17.1 \%  & 25.0 \%          & \textbf{22.0 \%}  \\ 
       & \textbf{Click}                        & \textbf{65.2 \%}  & 12.6 \%  & 37.3 \%          & 44.0 \%           \\  
       & \textbf{6 kHz}                        & 1.4 \%            & 22.0 \%  & \textbf{16.4 \%} & 14.3 \%           \\  
       & \textbf{12 kHz}                       & 1.4 \%            & 22.2 \%  & \textbf{25.8 \%} & 17.5 \%           \\  
       & \textbf{18 kHz}                       & 6.3 \%            & 15.1 \%  & \textbf{24.0 \%} & 18.7 \%           \\  
       & \textbf{24 kHz}                       & 12.1 \%           & 14.1 \%  & \textbf{23.8 \%} & 18.7 \%           \\ 
       & \textbf{30 kHz}                       & 20.2 \%           & 16.5 \%  & \textbf{22.9 \%} & 19.2 \%           \\ \hline

\\ 

\multicolumn{2}{l|}{\cellcolor{Lightgray}\textbf{\textit{B) $\pm$5 dB match}}}                      
                                  & \multicolumn{1}{c}{\textbf{experiment 1}} 
                                  & \multicolumn{1}{c|}{\textbf{experiment 3}} 
                                  & \multicolumn{1}{c}{\textbf{experiment 5}} 
                                  & \multicolumn{1}{c}{\textbf{experiment 7}}                                          \\ \hline
\multirow{7}{*}{\textbf{\begin{tabular}[r]{@{}r@{}}GMC\\data\end{tabular}}} 
       & \multicolumn{1}{r|}{\textbf{Overall}} & \textbf{75.6 \%}  & 35.7 \%   & 72.1 \%           & 73 \%             \\  
       & \multicolumn{1}{r|}{\textbf{Click}}   & 90.3 \%           & 58.3 \%   & \textbf{95.4 \%}  & 95.2 \%           \\  
       & \multicolumn{1}{r|}{\textbf{6 kHz}}   & \textbf{70.3 \%}  & 36.0 \%   & 58.7 \%           & 57.6 \%           \\  
       & \multicolumn{1}{r|}{\textbf{12 kHz}}  & \textbf{80.1 \%}  & 28.1 \%   & 66.4 \%           & 69.9 \%           \\  
       & \multicolumn{1}{r|}{\textbf{18 kHz}}  & \textbf{78.4 \%}  & 28.1 \%   & 69.9 \%           & 72.7 \%           \\  
       & \multicolumn{1}{r|}{\textbf{24 kHz}}  & \textbf{73.9 \%}  & 32.5 \%   & 73.3 \%           & 72.1 \%           \\ 
       & \multicolumn{1}{r|}{\textbf{30 kHz}}  & 60.5 \%           & 31.0 \%   & 69 \%             & \textbf{70.7 \%}  \\ \cline{3-6}
\multicolumn{2}{l|}{}             & \multicolumn{1}{c}{\textbf{experiment 2}} 
                                  & \multicolumn{1}{c|}{\textbf{experiment 4}} 
                                  & \multicolumn{1}{c}{\textbf{experiment 6}} 
                                  & \multicolumn{1}{c}{\textbf{experiment 8}}                                          \\ \cline{3-6}
\multirow{7}{*}{\textbf{\begin{tabular}[r]{@{}r@{}}ING\\data\end{tabular}}} 
       & \textbf{Overall}                      & 40.1 \%           & \textbf{77.7 \%}  & 66.3 \%    & 61.5 \%          \\  
       & \textbf{Click}                        & \textbf{98.3 \%}  & 83.9 \%           & 89.9 \%    & 91.2 \%          \\  
       & \textbf{6 kHz}                        & 3.0 \%            & \textbf{77.1 \%}  & 51.7 \%    & 48.9 \%          \\ 
       & \textbf{12 kHz}                       & 3.6 \%            & \textbf{79.5 \%}  & 63.4 \%    & 54.1 \%          \\  
       & \textbf{18 kHz}                       & 34.2 \%           & \textbf{75.3 \%}  & 62.8 \%    & 58.2 \%          \\ 
       & \textbf{24 kHz}                       & 47.0 \%           & \textbf{76.9 \%}  & 65.2 \%    & 58.5 \%          \\ 
       & \textbf{30 kHz}                       & 55.4 \%           & \textbf{73.6 \%}  & 65.1 \%    & 58.4 \%          \\ \hline

\\ 

\multicolumn{2}{r|}{\cellcolor{Lightgray}\textbf{\textit{C) $\pm$10 dB match}}}    
                                  & \multicolumn{1}{c}{\textbf{experiment 1}} 
                                  & \multicolumn{1}{c|}{\textbf{experiment 3}} 
                                  & \multicolumn{1}{c}{\textbf{experiment 5}} 
                                  & \multicolumn{1}{c}{\textbf{experiment 7}}                                                 \\ \hline
\multirow{7}{*}{\textbf{\begin{tabular}[r]{@{}r@{}}GMC\\data\end{tabular}}} 
       & \multicolumn{1}{r|}{\textbf{Overall}} & \textbf{89.8 \%}  & 62.3 \%           & 88.0 \%          & 88.2 \%           \\  
       & \multicolumn{1}{r|}{\textbf{Click}}   & 98.5 \%           & 85.4 \%           & 98.7 \%          & \textbf{98.9 \%}  \\  
       & \multicolumn{1}{r|}{\textbf{6 kHz}}   & \textbf{87.9 \%}  & 59.0 \%           & 79.8 \%          & 78.5 \%           \\  
       & \multicolumn{1}{r|}{\textbf{12 kHz}}  & \textbf{93.9 \%}  & 62.8 \%           & 85.9 \%          & 87.5 \%           \\  
       & \multicolumn{1}{r|}{\textbf{18 kHz}}  & \textbf{93.4 \%}  & 57.4 \%           & 89.7 \%          & 90.1 \%           \\ 
       & \multicolumn{1}{r|}{\textbf{24 kHz}}  & 88.0 \%           & 55.4 \%           & \textbf{89.3 \%} & 88.8 \%           \\ 
       & \multicolumn{1}{r|}{\textbf{30 kHz}}  & 77.4 \%           & 53.4 \%           & 84.7 \%          & \textbf{85.5 \%}  \\ \cline{3-6}
\multicolumn{2}{l|}{}             & \multicolumn{1}{c}{\textbf{experiment 2}} 
                                  & \multicolumn{1}{c|}{\textbf{experiment 4}} 
                                  & \multicolumn{1}{c}{\textbf{experiment 6}} 
                                  & \multicolumn{1}{c}{\textbf{experiment 8}}                                                 \\ \cline{3-6}
\multirow{7}{*}{\textbf{\begin{tabular}[r]{@{}r@{}}ING\\ data\end{tabular}}} 
       & \textbf{Overall}                      & 58.9 \%           & \textbf{96.3 \%}  & 85.9 \%          & 83.9 \%           \\  
       & \textbf{Click}                        & \textbf{99.7 \%}  & 99.3 \%           & 98.1 \%          & 98.3 \%           \\  
       & \textbf{6 kHz}                        & 8.6 \%            & \textbf{94.9 \%}  & 75.0 \%          & 74.5 \%           \\ 
       & \textbf{12 kHz}                       & 14.6 \%           & \textbf{95.9 \%}  & 83.8 \%          & 79.6 \%           \\  
       & \textbf{18 kHz}                       & 71.0 \%           & \textbf{96.0 \%}  & 85.3 \%          & 83.6 \%           \\ 
       & \textbf{24 kHz}                       & 79.5 \%           & \textbf{96.6 \%}  & 87.3 \%          & 84.2 \%           \\ 
       & \textbf{30 kHz}                       & 80.5 \%           & \textbf{95.4 \%}  & 86.0 \%          & 83.2 \%  
\end{tabular}
\end{small}
\end{table}

%% file: supplements/supplement1.tex
\subsection*{Model I}

Using a 1000 time step input vector of any stimulus frequency, this model predicts:
\begin{itemize}
\item response yes/no (0/1)
\item the frequency of the stimulus (click, 6, 12, 18, 24, 30 kHz)
\item the sound level of the stimulus (5, 10, ..., 95 dB)
\end{itemize}

\begin{scriptsize}
\begin{verbatim}
__________________________________________________________________________________________________
Layer (type)                    Output Shape         Param #     Connected to                     
==================================================================================================
input_3 (InputLayer)            (None, 1000, 1)      0                                            
__________________________________________________________________________________________________
batch_normalization_v1_13 (Batc (None, 1000, 1)      4           input_3[0][0]                    
__________________________________________________________________________________________________
conv1d_11 (Conv1D)              (None, 1000, 256)    65792       batch_normalization_v1_13[0][0]  
__________________________________________________________________________________________________
batch_normalization_v1_14 (Batc (None, 1000, 256)    1024        conv1d_11[0][0]                  
__________________________________________________________________________________________________
activation_11 (Activation)      (None, 1000, 256)    0           batch_normalization_v1_14[0][0]  
__________________________________________________________________________________________________
conv1d_12 (Conv1D)              (None, 1000, 128)    4194432     activation_11[0][0]              
__________________________________________________________________________________________________
batch_normalization_v1_15 (Batc (None, 1000, 128)    512         conv1d_12[0][0]                  
__________________________________________________________________________________________________
activation_12 (Activation)      (None, 1000, 128)    0           batch_normalization_v1_15[0][0]  
__________________________________________________________________________________________________
average_pooling1d (AveragePooli (None, 250, 128)     0           activation_12[0][0]              
__________________________________________________________________________________________________
dropout_5 (Dropout)             (None, 250, 128)     0           average_pooling1d[0][0]          
__________________________________________________________________________________________________
conv1d_13 (Conv1D)              (None, 250, 64)      524352      dropout_5[0][0]                  
__________________________________________________________________________________________________
batch_normalization_v1_16 (Batc (None, 250, 64)      256         conv1d_13[0][0]                  
__________________________________________________________________________________________________
activation_13 (Activation)      (None, 250, 64)      0           batch_normalization_v1_16[0][0]  
__________________________________________________________________________________________________
conv1d_14 (Conv1D)              (None, 250, 32)      65568       activation_13[0][0]              
__________________________________________________________________________________________________
batch_normalization_v1_17 (Batc (None, 250, 32)      128         conv1d_14[0][0]                  
__________________________________________________________________________________________________
activation_14 (Activation)      (None, 250, 32)      0           batch_normalization_v1_17[0][0]  
__________________________________________________________________________________________________
average_pooling1d_1 (AveragePoo (None, 125, 32)      0           activation_14[0][0]              
__________________________________________________________________________________________________
dropout_6 (Dropout)             (None, 125, 32)      0           average_pooling1d_1[0][0]        
__________________________________________________________________________________________________
conv1d_15 (Conv1D)              (None, 125, 16)      8208        dropout_6[0][0]                  
__________________________________________________________________________________________________
batch_normalization_v1_18 (Batc (None, 125, 16)      64          conv1d_15[0][0]                  
__________________________________________________________________________________________________
activation_15 (Activation)      (None, 125, 16)      0           batch_normalization_v1_18[0][0]  
__________________________________________________________________________________________________
conv1d_16 (Conv1D)              (None, 125, 8)       1032        activation_15[0][0]              
__________________________________________________________________________________________________
batch_normalization_v1_19 (Batc (None, 125, 8)       32          conv1d_16[0][0]                  
__________________________________________________________________________________________________
activation_16 (Activation)      (None, 125, 8)       0           batch_normalization_v1_19[0][0]  
__________________________________________________________________________________________________
average_pooling1d_2 (AveragePoo (None, 62, 8)        0           activation_16[0][0]              
__________________________________________________________________________________________________
dropout_7 (Dropout)             (None, 62, 8)        0           average_pooling1d_2[0][0]        
__________________________________________________________________________________________________
conv1d_17 (Conv1D)              (None, 62, 4)        132         dropout_7[0][0]                  
__________________________________________________________________________________________________
batch_normalization_v1_20 (Batc (None, 62, 4)        16          conv1d_17[0][0]                  
__________________________________________________________________________________________________
activation_17 (Activation)      (None, 62, 4)        0           batch_normalization_v1_20[0][0]  
__________________________________________________________________________________________________
conv1d_18 (Conv1D)              (None, 62, 1)        9           activation_17[0][0]              
__________________________________________________________________________________________________
batch_normalization_v1_21 (Batc (None, 62, 1)        4           conv1d_18[0][0]                  
__________________________________________________________________________________________________
activation_18 (Activation)      (None, 62, 1)        0           batch_normalization_v1_21[0][0]  
__________________________________________________________________________________________________
flatten_2 (Flatten)             (None, 62)           0           activation_18[0][0]              
__________________________________________________________________________________________________
dense_2 (Dense)                 (None, 32)           2016        flatten_2[0][0]                  
__________________________________________________________________________________________________
main_prediction (Dense)         (None, 1)            33          dense_2[0][0]                    
__________________________________________________________________________________________________
frequency_prediction (Dense)    (None, 6)            198         dense_2[0][0]                    
__________________________________________________________________________________________________
sl_prediction (Dense)           (None, 20)           660         dense_2[0][0]                    
==================================================================================================
Total params: 4,864,472
Trainable params: 4,863,452
Non-trainable params: 1,020
\end{verbatim}
\end{scriptsize}

\subsection*{Model II}

Using an input vector of 20 sound level prediction scores from model I output, this model predicts:
\begin{itemize}
\item the frequency of the stimulus (click, 6, 12, 18, 24, 30 kHz)
\item the hearing threshold (5, 10, ..., 95 dB)
\end{itemize}

\begin{scriptsize}
\begin{verbatim}
__________________________________________________________________________________________________
Layer (type)                    Output Shape         Param #     Connected to                     
==================================================================================================
input_2 (InputLayer)            (None, 20, 1)        0                                            
__________________________________________________________________________________________________
batch_normalization_v1_4 (Batch (None, 20, 1)        4           input_2[0][0]                    
__________________________________________________________________________________________________
conv1d_3 (Conv1D)               (None, 20, 128)      896         batch_normalization_v1_4[0][0]   
__________________________________________________________________________________________________
batch_normalization_v1_5 (Batch (None, 20, 128)      512         conv1d_3[0][0]                   
__________________________________________________________________________________________________
activation_3 (Activation)       (None, 20, 128)      0           batch_normalization_v1_5[0][0]   
__________________________________________________________________________________________________
conv1d_4 (Conv1D)               (None, 20, 64)       41024       activation_3[0][0]               
__________________________________________________________________________________________________
batch_normalization_v1_6 (Batch (None, 20, 64)       256         conv1d_4[0][0]                   
__________________________________________________________________________________________________
activation_4 (Activation)       (None, 20, 64)       0           batch_normalization_v1_6[0][0]   
__________________________________________________________________________________________________
max_pooling1d_1 (MaxPooling1D)  (None, 6, 64)        0           activation_4[0][0]               
__________________________________________________________________________________________________
dropout_2 (Dropout)             (None, 6, 64)        0           max_pooling1d_1[0][0]            
__________________________________________________________________________________________________
conv1d_5 (Conv1D)               (None, 6, 32)        8224        dropout_2[0][0]                  
__________________________________________________________________________________________________
batch_normalization_v1_7 (Batch (None, 6, 32)        128         conv1d_5[0][0]                   
__________________________________________________________________________________________________
activation_5 (Activation)       (None, 6, 32)        0           batch_normalization_v1_7[0][0]   
__________________________________________________________________________________________________
conv1d_6 (Conv1D)               (None, 6, 16)        1552        activation_5[0][0]               
__________________________________________________________________________________________________
batch_normalization_v1_8 (Batch (None, 6, 16)        64          conv1d_6[0][0]                   
__________________________________________________________________________________________________
activation_6 (Activation)       (None, 6, 16)        0           batch_normalization_v1_8[0][0]   
__________________________________________________________________________________________________
flatten_1 (Flatten)             (None, 96)           0           activation_6[0][0]               
__________________________________________________________________________________________________
dense_1 (Dense)                 (None, 64)           6208        flatten_1[0][0]                  
__________________________________________________________________________________________________
main_prediction (Dense)         (None, 21)           1365        dense_1[0][0]                    
__________________________________________________________________________________________________
frequency_prediction (Dense)    (None, 6)            390         dense_1[0][0]                    
==================================================================================================
Total params: 60,623
Trainable params: 60,141
Non-trainable params: 482
\end{verbatim}
\end{scriptsize}

%% file: supplements/supplement2.tex
\begin{figure}[H]
\centering
\subfloat[NN/SLR GMC-ING (experiments 2 and 6)]{%
  \includegraphics[width=0.9\linewidth]{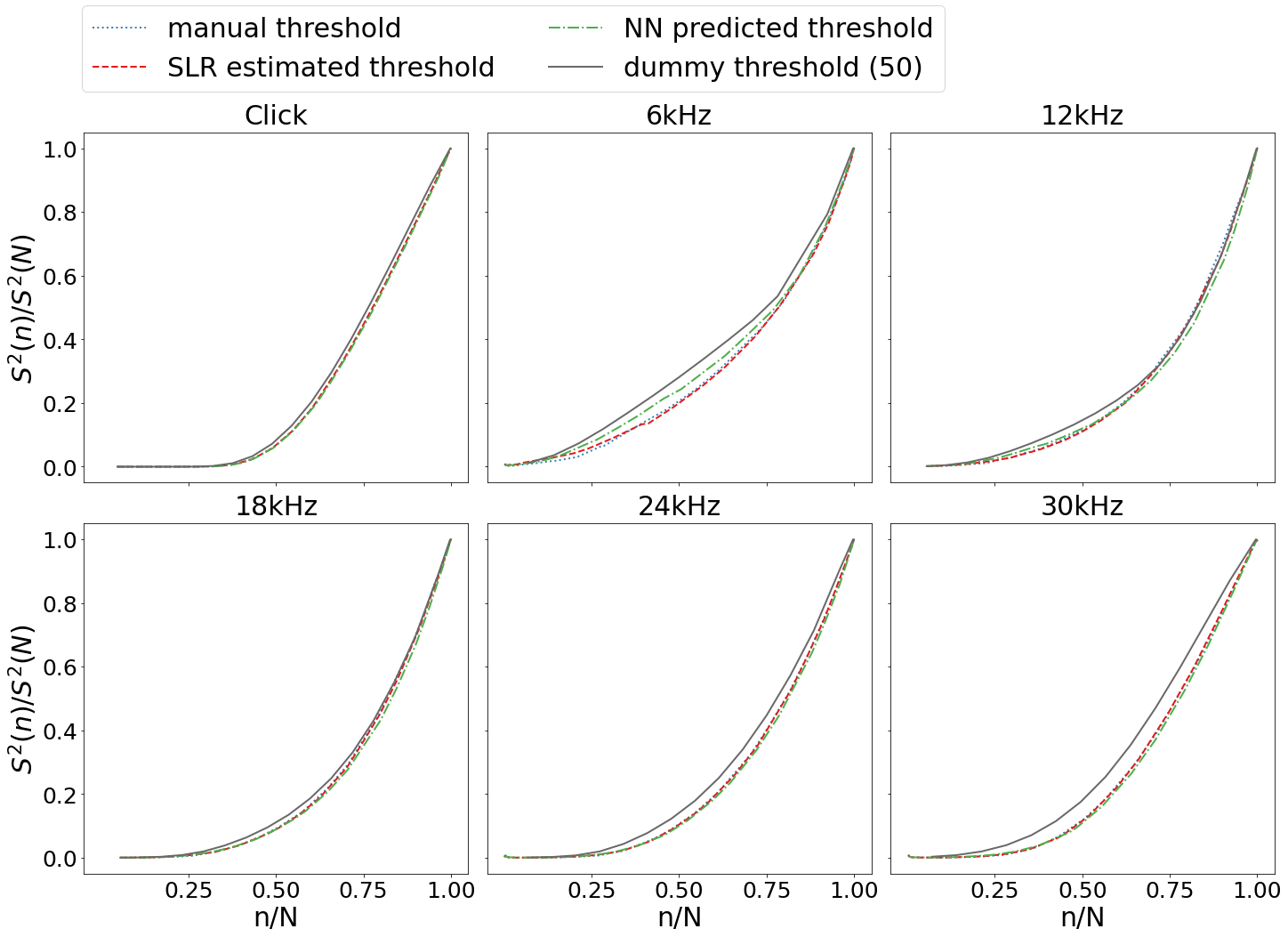}
  \label{fig:evaluation_curves-exp2_6-GMC-ING}
  }
  
\subfloat[NN/SLR ING-GMC (experiments 3 and 7)]{%
  \includegraphics[width=0.9\linewidth]{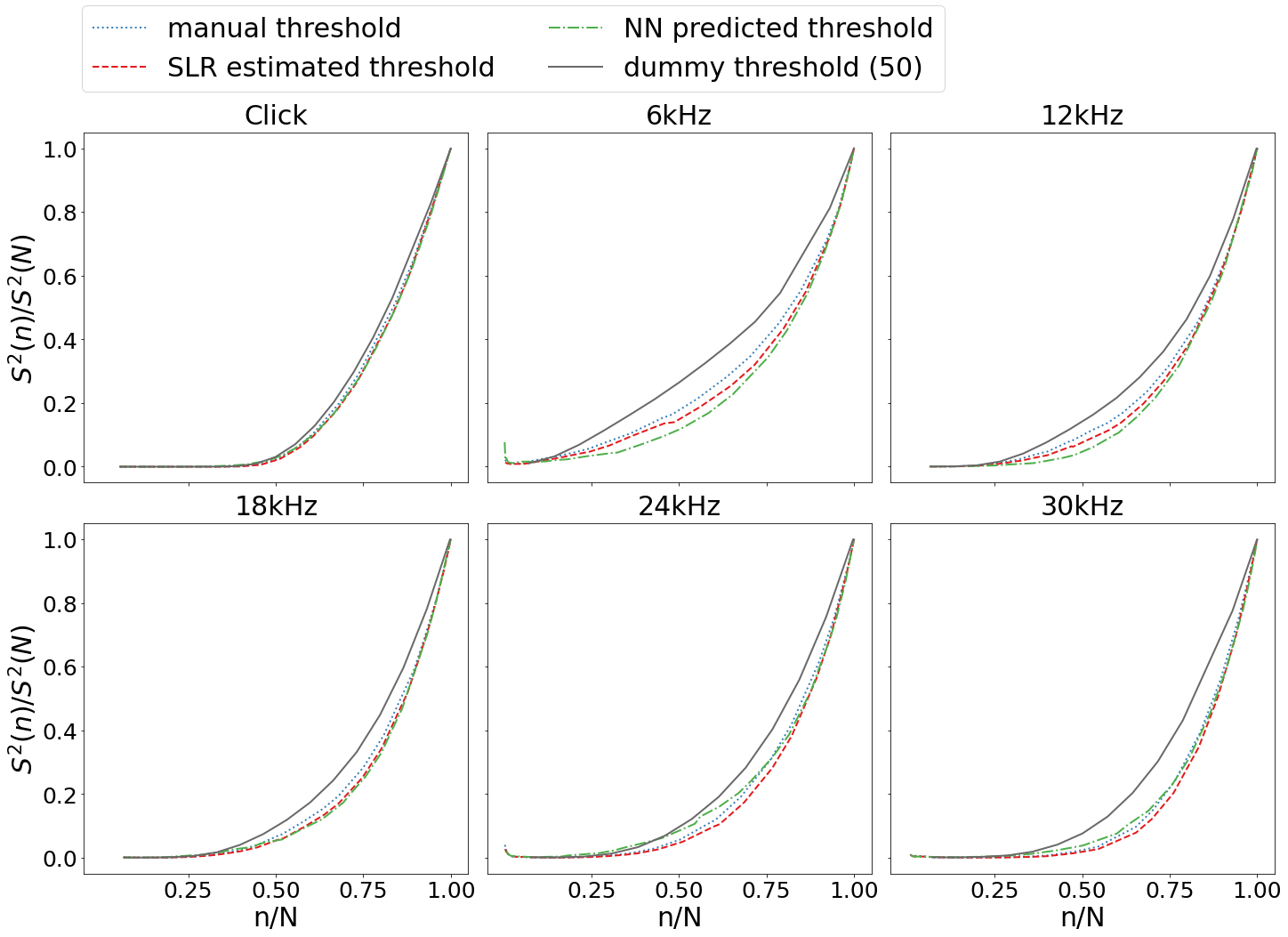}
  \label{fig:evaluation_curves-exp3_7-ING-GMC}
  }
\thisfloatpagestyle{empty}  
\caption{\textbf{Objective comparison of the quality of threshold finding methods using evaluation curves}. Evaluation curves allow the relative comparison of threshold finding methods without requiring absolute ground truth labels. Here, four methods are compared: manual thresholds (blue, dotted lines), SLR estimations (red, dashed lines), NN predictions (green, dash-dotted lines), and a ``always 50 dB'' dummy method (grey, solid lines). }
\label{fig:sup_eval_curves}
\end{figure}

\begin{figure}[H]
    \ContinuedFloat
    \caption[]{(continued from previous page)\\[0.2em]
    Separate plots show evaluation curves for each stimulus (click, 6 kHz - 30 kHz). Plots show the normalized time variance of the averaged signal $S^2(n)/S^2(N)$ (y-axis) vs.  the total percentage of ABR curves included in the cumulative average $n/N$ (x-axis). a) shows NN predictions and SLR estimations from experiments 2 and 6, b) shows NN predictions and SLR estimations from experiments 3 and 7, as introduced in \mbox{table \ref{tab:8experiments}}. 
    Two methods can be compared in a way that the evaluation curve of the better method deviates from zero later. Ideally, the curve of the best method is always below all other curves. Strangely, when GMC-trained/calibrated models where tested on ING data (experiments 2 and 6), human, NN and SLR were not much better than the dummy method. In addition, except for the 6 kHz stimulus, they did not differ much from each other. In contrast, when ING-trained/calibrated models were tested on GMC data (experiments 3 and 7), NN and SLR models were mostly better than the human method, with NN being best for click and 6 kHz to 18 kHz and SLR being best for 24 kHz and 30 kHz. }
\end{figure}

%% file: supplements/supplement3.tex
\begin{figure}[H]
\centering
\includegraphics[width=0.9\linewidth]{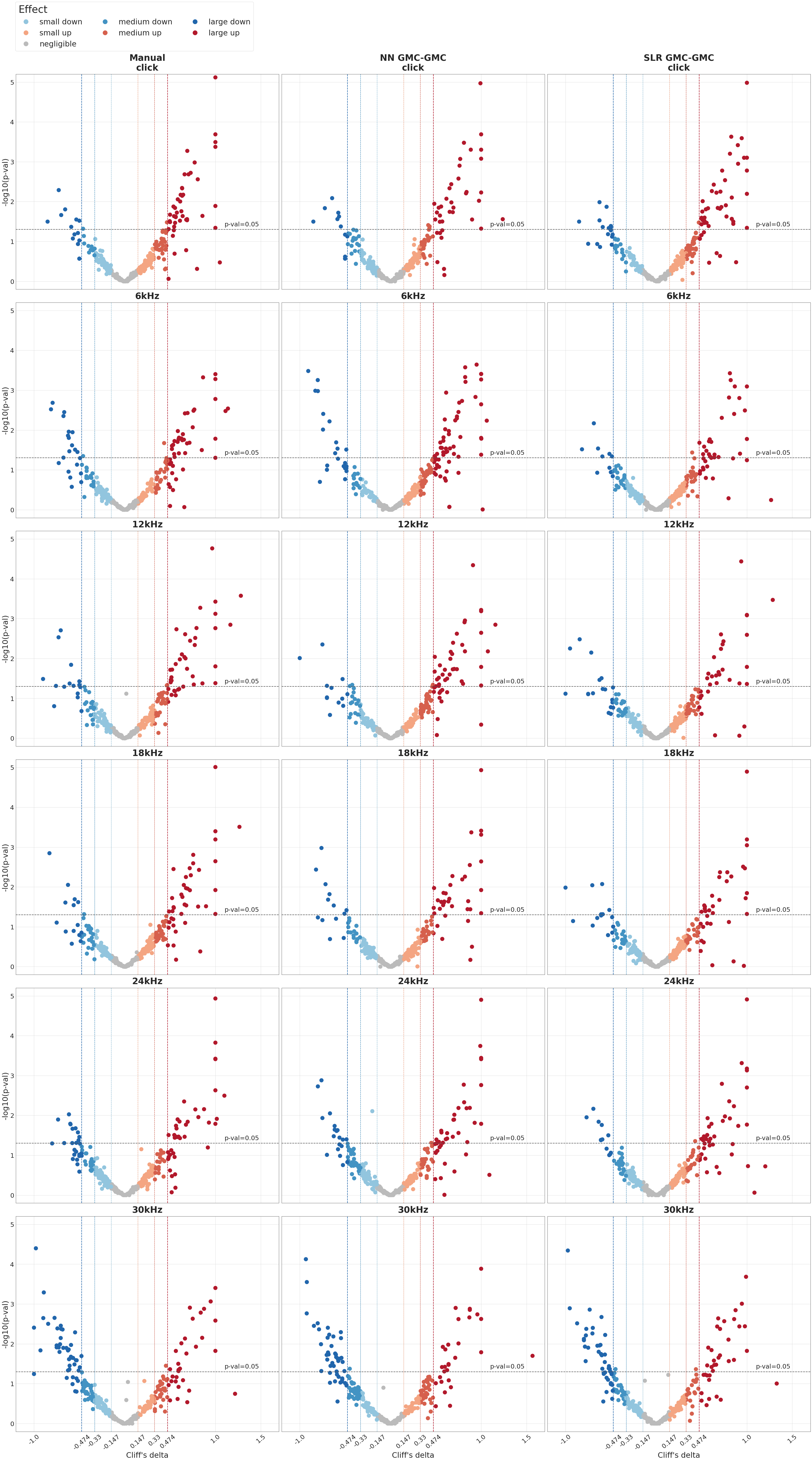}
\thisfloatpagestyle{empty}  
\caption{}
\label{fig:GMC_volcano_plot}
\end{figure}

\begin{figure}[H]
    \ContinuedFloat
    \caption[]{(continued from previous page)\\[0.2em]
    \textbf{Biologically relevant changes in hearing thresholds - GMC lines, all stimuli}. For each GMC line, Volcano plots show significance vs. relevance for comparisons of mutant and control mice. For each mouse line, represented by a dot, hearing thresholds were used to calculate significance (Wilcoxon test, y-axis) and non-parametric effect size (Cliff's delta \cite{cliff1993}, x-axis) of mutant vs. control animals.  Vertical lines indicate margins for small (0.147), medium (0.33) and large (0.474) effects as suggested in \cite{romano2006}. The horizontal line indicates the 0.05 significance threshold level. Accordingly, mutant lines represented by data points in the upper left and upper right areas denote lines with significant as well as relevant changes and thus are considered worthwhile candidates (see supplement table \ref{tab:top_candidates_click}). Dot colors in addition represent effect size as shown in the legend. Plot rows represent different stimuli (click, 6 kHz - 30 kHz). Columns compare the three hearing threshold finding methods compared in this work (left: manual, middle: NN, right: SLR).}
\end{figure}

%% file: supplements/supplement4.tex
\begin{table}[]
\centering
\caption{\textbf{Comparison of top candidate genes with modified click hearing threshold}. The columns show genes with a strong effect size for mutant-control group comparisons of click hearing thresholds. In each column, genes are ordered by non-parametric effect size (Cliff's delta \cite{cliff1993}), in descending order. Only genes with a significant ($p < 0.05$) and large effect size ($|d| > 0.474$, \cite{romano2006}) in any of the three methods are shown. Column headers indicate the direction of the effect and the method that was used to identify the hearing thresholds, respectively.  The ``combined'' column contains the combined list of all three methods.}
\label{tab:top_candidates_click}
\begin{small}
\begin{tabular}{$l^l^l^l|^l^l^l^l}
\multicolumn{4}{c|}{\textbf{increased click threshold}}  & \multicolumn{4}{c}{\textbf{decreased click threshold}}     \\ 
\textbf{combined} & \textbf{manual} & \textbf{NN} & \textbf{SLR} & \textbf{combined} & \textbf{manual} & \textbf{NN} & \textbf{SLR} \\ \hline
\rowstyle{\itshape}Palm3               & Zfp280d           & Slc20a2       & Lsm1           & Dio1                  & Dio1                & Dio1            & Dio1             \\
\rowstyle{\itshape}Zfp280d             & Lsm1              & Prkd2         & Strbp          & Gstm6                 & Gstm6               & Ucp1            & Hnf4a            \\
\rowstyle{\itshape}Strbp               & Strbp             & Zfp280d       & Palm3          & Hepacam2              & Hepacam2            & Cilp2           & Gstt1            \\
\rowstyle{\itshape}Hipk3               & Hipk3             & Strbp         & Zfp280d        & Slc25a15              & Slc25a15            & Gstm6           & Ostf1            \\
\rowstyle{\itshape}Prkd2               & Palm3             & Lsm1          & Hipk3          & Ostf1                 & Ostf1               & Raet1c          & Angptl3          \\
\rowstyle{\itshape}Lsm1                & Prkd2             & Palm3         & Prkd2          & Ucp1                  & Rab35               & Rab35           & Phactr4          \\
\rowstyle{\itshape}Mipol1              & Mipol1            & Hipk3         & Plag1          & Rab35                 & Cilp2               & Slc25a15        & Rab35            \\
\rowstyle{\itshape}Hoxa2               & Hoxa2             & Mipol1        & Hoxa2          & Hnf4a                 &                     &                 & Cilp2            \\
\rowstyle{\itshape}Slc20a2             & Plag1             & Vps13c        & Aspa           & Cilp2                 &                     &                 & Prox2            \\
\rowstyle{\itshape}Plag1               & Ldlr              & Plag1         & Dpp3           & Prox2                 &                     &                 &                  \\
\rowstyle{\itshape}Ldlr                & Hunk              & Nacc1         & Vps13c         & Phactr4               &                     &                 &                  \\
\rowstyle{\itshape}Hunk                & Atp5g2            & Btbd9         & Mipol1         & Raet1c                &                     &                 &                  \\
\rowstyle{\itshape}Atp5g2              & Ube3c             & Alkbh6        & Btbd9          & Gstt1                 &                     &                 &                  \\
\rowstyle{\itshape}Btbd9               & Btbd9             & Zdhhc5        & Alkbh6         & Angptl3               &                     &                 &                  \\
\rowstyle{\itshape}Ube3c               & Nacc1             & Rabgap1       & Zdhhc5         &                       &                     &                 &                  \\
\rowstyle{\itshape}Nacc1               & Tle1              & Aspa          & Bms1           &                       &                     &                 &                  \\
\rowstyle{\itshape}Tle1                & Gpatch2l          & Hoxa2         & Nacc1          &                       &                     &                 &                  \\
\rowstyle{\itshape}Gpatch2l            & Cidec             & Sytl4         & Plekha1        &                       &                     &                 &                  \\
\rowstyle{\itshape}Cidec               & Alkbh6            & Pdcd5         & Ldlr           &                       &                     &                 &                  \\
\rowstyle{\itshape}Alkbh6              & Spryd3            & Ldlr          & Adprm          &                       &                     &                 &                  \\
\rowstyle{\itshape}Spryd3              & Bccip             & Ttll12        & Wrnip1         &                       &                     &                 &                  \\
\rowstyle{\itshape}Bccip               & Pdcd5             & Hunk          & Sytl4          &                       &                     &                 &                  \\
\rowstyle{\itshape}Pdcd5               & Ppp4r3b           & Ppp4r3b       & Hdac1          &                       &                     &                 &                  \\
\rowstyle{\itshape}Ppp4r3b             & Csnk1g2           & Kansl1l       & Ppp4r3b        &                       &                     &                 &                  \\
\rowstyle{\itshape}Csnk1g2             & Ngdn              & Bccip         & Tle1           &                       &                     &                 &                  \\
\rowstyle{\itshape}Ngdn                & Gpsm2             & Csnk1g2       & Gpsm2          &                       &                     &                 &                  \\
\rowstyle{\itshape}Gpsm2               & Aspa              & Plekha1       & Pkn2           &                       &                     &                 &                  \\
\rowstyle{\itshape}Aspa                & Plekha1           & Tle1          & Me2            &                       &                     &                 &                  \\
\rowstyle{\itshape}Pfkfb3              & Pfkfb3            & Fdx1          & Ttll12         &                       &                     &                 &                  \\
\rowstyle{\itshape}Plekha1             & Wrnip1            & Gpsm2         & Rabgap1        &                       &                     &                 &                  \\
\rowstyle{\itshape}Wrnip1              & Zdhhc5            & Atp5g2        & Csnk1g2        &                       &                     &                 &                  \\
\rowstyle{\itshape}Zdhhc5              & Uggt2             &               & Tbl1xr1        &                       &                     &                 &                  \\
\rowstyle{\itshape}Dpp3                & Gdi2              &               & Pdcd5          &                       &                     &                 &                  \\
\rowstyle{\itshape}Uggt2               & Sec14l4           &               & Hunk           &                       &                     &                 &                  \\
\rowstyle{\itshape}Gdi2                & Tanc2             &               & Atp5g2         &                       &                     &                 &                  \\
\rowstyle{\itshape}Sec14l4             & Rfxank            &               & Fdx1           &                       &                     &                 &                  \\
\rowstyle{\itshape}Sytl4               & Me2               &               & Tanc2          &                       &                     &                 &                  \\
\rowstyle{\itshape}Rfxank              & Gsk3a             &               & Cenpv          &                       &                     &                 &                  \\
\rowstyle{\itshape}Tanc2               &                   &               & Bccip          &                       &                     &                 &                  \\
\rowstyle{\itshape}Me2                 &                   &               &                &                       &                     &                 &                  \\
\rowstyle{\itshape}Gsk3a               &                   &               &                &                       &                     &                 &                  \\
\rowstyle{\itshape}Rabgap1             &                   &               &                &                       &                     &                 &                  \\
\rowstyle{\itshape}Ttll12              &                   &               &                &                       &                     &                 &                  \\
\rowstyle{\itshape}Fdx1                &                   &               &                &                       &                     &                 &                  \\
\rowstyle{\itshape}Bms1                &                   &               &                &                       &                     &                 &                  \\
\rowstyle{\itshape}Hdac1               &                   &               &                &                       &                     &                 &                  \\
\rowstyle{\itshape}Tbl1xr1             &                   &               &                &                       &                     &                 &                  \\
\rowstyle{\itshape}Adprm               &                   &               &                &                       &                     &                 &                  \\
\rowstyle{\itshape}Vps13c              &                   &               &                &                       &                     &                 &                  \\
\rowstyle{\itshape}Kansl1l             &                   &               &                &                       &                     &                 &                  \\
\rowstyle{\itshape}Pkn2                &                   &               &                &                       &                     &                 &                  \\
\rowstyle{\itshape}Cenpv               &                   &               &                &                       &                     &                 &                 
\end{tabular}
\end{small}
\end{table}

%% file: supplements/supplement5.tex
\begin{table}[]
\centering
\caption{\textbf{Comparison of top candidate genes with modified 30 kHz hearing threshold}. The columns show genes with a strong effect size for mutant-control group comparisons of 30 kHz hearing thresholds. In each column, genes are ordered by non-parametric effect size (Cliff's delta \cite{cliff1993}), in descending order. Only genes with a significant ($p < 0.05$) and large effect size ($|d| > 0.474$, \cite{romano2006}) in any of the three methods are shown. Column headers indicate the direction of the effect and the method that was used to identify the hearing thresholds, respectively. The ``combined'' column contains the ranked combined list of all three methods.}
\label{tab:top_candidates_30kHz}
\begin{small}
\begin{tabular}{$l^l^l^l|^l^l^l^l}
\multicolumn{4}{c|}{\textbf{increased 30 kHz threshold}}  & \multicolumn{4}{c}{\textbf{decreased 30 kHz threshold}}     \\ 
\textbf{combined} & \textbf{manual} & \textbf{NN} & \textbf{SLR} & \textbf{combined} & \textbf{manual} & \textbf{NN} & \textbf{SLR} \\ \hline
\rowstyle{\itshape}Chst5   & Zfp280d & Chst5   & Zfp280d & Miga1    & Miga1   & Gfpt2    & Gfpt2   \\
\rowstyle{\itshape}Palm3   & Strbp   & Zfp280d & Plag1   & Gfpt2    & Gfpt2   & Rest     & Hnf4a   \\
\rowstyle{\itshape}Strbp   & Palm3   & Strbp   & Strbp   & Aqp6     & Aqp6    & Tap1     & Rest    \\
\rowstyle{\itshape}Zfp280d & Uggt2   & Plag1   & Ttll12  & Rest     & Rest    & Hnf4a    & Fgfr1op \\
\rowstyle{\itshape}Uggt2   & Prkd2   & Ube3c   & Ube3c   & Cotl1    & Cotl1   & Becn1    & Msh5    \\
\rowstyle{\itshape}Prkd2   & Ube3c   & Prkd2   & Prkd2   & Hnf4a    & Hnf4a   & Fbp2     & Dis3    \\
\rowstyle{\itshape}Ube3c   & Plag1   & Palm3   & Palm3   & Cenph    & Cenph   & Msh5     & Miga1   \\
\rowstyle{\itshape}Plag1   & Nacc1   & Uggt2   & Mipol1  & Dnajc27  & Msh5    & Aqp6     & Fbp2    \\
\rowstyle{\itshape}Nacc1   & Gsk3a   & Ttll12  & Mthfsl  & Msh5     & Dnajc27 & Etfdh    & Becn1   \\
\rowstyle{\itshape}Gsk3a   & Hipk3   & Gsk3a   & Cnot6l  & Lama1    & Lama1   & Ostf1    & Dnajc27 \\
\rowstyle{\itshape}Hipk3   & Ttll12  & Hipk3   & Gsk3a   & Pkig     & Pkig    & Cilp2    & Ppy     \\
\rowstyle{\itshape}Ttll12  & Zdhhc5  & Nacc1   & Csnk1g2 & Fgfr1op  & Fgfr1op & Lss      & Tap1    \\
\rowstyle{\itshape}Zdhhc5  & Ldlr    & Rnf186  & Rnf186  & Fbp2     & Fbp2    & Gstm6    & Fam162a \\
\rowstyle{\itshape}Mipol1  & Ngdn    & Adprm   & Adprm   & Cyp7a1   & Cyp7a1  & Pkig     & Cenph   \\
\rowstyle{\itshape}Ldlr    & Adprm   & Zdhhc5  & Hipk3   & Fam162a  & Fam162a & Galk2    & Cyp7a1  \\
\rowstyle{\itshape}Ngdn    & Pdcd5   & Aldh1l1 & Alkbh6  & Galk2    & Galk2   & Mgat1    & Galk2   \\
\rowstyle{\itshape}Adprm   & Rnf186  & Csnk1g2 & Aldh1l1 & Yae1d1   & Yae1d1  & Slc25a15 & Lss     \\
\rowstyle{\itshape}Rnf186  & Fabp2   & Paox    & Pdcd5   & Becn1    & Becn1   & Ppy      & Cotl1   \\
\rowstyle{\itshape}Pdcd5   & Entpd1  & Cnot6l  & Uggt2   & Ly6g6d   & Ly6g6d  & Ctc1     & Pdhx    \\
\rowstyle{\itshape}Fabp2   & Csnk1g2 & Alkbh6  & Nacc1   & Jmjd6    & Jmjd6   & Sarnp    & Vwa8    \\
\rowstyle{\itshape}Aldh1l1 &         & Fabp2   & Paox    & Vwa8     & Dis3    & Vwa8     & Cilp2   \\
\rowstyle{\itshape}Entpd1  &         &         & Zdhhc5  & Dis3     & Vwa8    & Cyp7a1   & Zbtb24  \\
\rowstyle{\itshape}Cnot6l  &         &         & Gpsm2   & Tap1     & Tap1    & Dnajc27  & Mgat1   \\
\rowstyle{\itshape}Csnk1g2 &         &         & Npepps  & Phactr4  & Phactr4 & Fgfr1op  & Yae1d1  \\
\rowstyle{\itshape}Paox    &         &         &         & C1galt1  & C1galt1 & Nubp2    & Lactb   \\
\rowstyle{\itshape}Mthfsl  &         &         &         & Cth      & Cth     & Cotl1    & Sarnp   \\
\rowstyle{\itshape}Gpsm2   &         &         &         & Cilp2    & Cilp2   & Fam162a  & Scmh1   \\
\rowstyle{\itshape}Npepps  &         &         &         & Ppy      & Ppy     & Pdhx     & Rcn1    \\
\rowstyle{\itshape}Alkbh6  &         &         &         & Cdc123   & Cdc123  & Lactb    & Nubp2   \\
\rowstyle{\itshape}        &         &         &         & Scmh1    & Scmh1   & Capn12   &         \\
\rowstyle{\itshape}        &         &         &         & Ndufa10  & Ndufa10 & Yae1d1   &         \\
\rowstyle{\itshape}        &         &         &         & Pdhx     & Pdhx    & Jmjd6    &         \\
\rowstyle{\itshape}        &         &         &         & Lss      & Lss     &          &         \\
\rowstyle{\itshape}        &         &         &         & Capn12   & Capn12  &          &         \\
\rowstyle{\itshape}        &         &         &         & Sarnp    & Sarnp   &          &         \\
\rowstyle{\itshape}        &         &         &         & Cystm1   & Cystm1  &          &         \\
\rowstyle{\itshape}        &         &         &         & Ube3a    & Ube3a   &          &         \\
\rowstyle{\itshape}        &         &         &         & Nubp2    &         &          &         \\
\rowstyle{\itshape}        &         &         &         & Slc25a15 &         &          &         \\
\rowstyle{\itshape}        &         &         &         & Zbtb24   &         &          &         \\
\rowstyle{\itshape}        &         &         &         & Rcn1     &         &          &         \\
\rowstyle{\itshape}        &         &         &         & Etfdh    &         &          &         \\
\rowstyle{\itshape}        &         &         &         & Ostf1    &         &          &         \\
\rowstyle{\itshape}        &         &         &         & Lactb    &         &          &         \\
\rowstyle{\itshape}        &         &         &         & Ctc1     &         &          &         \\
\rowstyle{\itshape}        &         &         &         & Mgat1    &         &          &         \\
\rowstyle{\itshape}        &         &         &         & Gstm6    &         &          &         
\end{tabular}
\end{small}
\end{table}

%% file: supplements/supplement6.tex
\texttt{Supplement\_SLR\_candidate\_genes.xlsx} contains basic information on the 76 candidate genes with a significant ($p < 0.05$) and large effect size ($|d| > 0.474$, \cite{romano2006}) between mutants and controls in click and/or 30 kHz threshold derived from SLR based analysis. Descriptions of the spreadsheet columns are given in the following table:

\begin{table}[h]
\begin{tabular}{c|l}
\textbf{column} & \textbf{description} \\ \hline
\textbf{A} & official gene symbol \\
\textbf{B} & is the gene mentioned as hearing related at https://hereditaryhearingloss.org - yes/no? \\
\textbf{C} & is the human orthologue linked to a human disease according to OMIM? If yes, which?     \\
\textbf{D} & OMIM ID(s)                                                                              \\
\textbf{E} & gene expression in inner ear according to \cite{ranum2019}                              \\
\textbf{F} & gene expression in inner ear according to \cite{scheffer2015}                           \\
\textbf{G} & mouse model, other than IMPC                                                            \\
\textbf{H} & threshold change at click stimulation - increase/decrease                               \\
\textbf{I} & threshold change at 30 kHz stimulation - increase/decrease                              \\
\textbf{J} & effect size at click stimulation                                                        \\
\textbf{K} & effect size at 30 kHz stimulation                                                       \\
\end{tabular}
\end{table}